\documentclass[12pt]{iopart}
\usepackage{amsfonts}
\usepackage{psfrag}
\usepackage{subfigure}
\usepackage{graphicx}
\eqnobysec
\usepackage{iopams}
\def\strutdepth{\dp\strutbox}
\def\nw#1{\strut\vadjust{\kern-\strutdepth\vtop to0pt{\vss\hbox to\hsize
{\hskip\hsize\hskip5pt$\leftarrow$\hss\strut}}}{\em #1}}

\begin{document}

\title[Singularities]{The role of self-similarity in singularities of PDE's}
\author{Jens Eggers$^*$ and Marco A. Fontelos$^\dagger$}

\begin{abstract}
We survey rigorous, formal, and numerical results on the formation of
point-like singularities (or blow-up) for a wide range of evolution
equations. We use a similarity transformation of the original equation with
respect to the blow-up point, such that self-similar behaviour is mapped to
the fixed point of a \textit{dynamical system}. We point out that analysing 
the dynamics close to the fixed point is a useful way of characterising 
the singularity, in that the dynamics frequently reduces to very few 
dimensions. As far as we are aware,
examples from the literature either correspond to stable fixed points,
low-dimensional centre-manifold dynamics, limit cycles, or travelling waves.
For each ``class'' of singularity, we give detailed examples.
\end{abstract}

\address{
$^*$School of Mathematics,
University of Bristol, University Walk, \\
Bristol BS8 1TW, United Kingdom \\
$^\dagger$  Instituto de Ciencias Matem\'aticas,
 (ICMAT, CSIC-UAM-UCM-UC3M), \\ C/ Serrano 123, 28006 
Madrid, Spain
          }
\maketitle

\section{Introduction}
Non-linear partial differential equations (PDE's) are distinguished by the
fact that, starting from smooth initial data, they can develop a singularity
in finite time \cite{Levine90,CP93,K97,Straughan98}. \footnote{Of course, 
there are also many examples of 
nonlinear PDE's for which global existence can be established!}
Very often, such a singularity
corresponds to a physical event, such as the solution (e.g. a physical flow
field) changing topology, and/or the emergence of a new (singular) 
structure, such as a tip, cusp, sheet, or jet. On the other hand, 
a singularity can also imply that some essential physics is missing from
the equation in question, which should thus be supplemented with 
additional terms. (Even in the latter case, the singularity may still
be indicative of a real physical event). 

Consider for
example the physical case shown in Fig.~\ref{lava}, which we will treat in
section \ref{travel} below. Shown is a snapshot of one viscous fluid
dripping into another fluid, close to the point where a drop of the inner
fluid pinches off. This process is driven by surface tension, which tries to
minimise the surface area between the two fluids. At a particular point 
$x_0,t_0$ in space and time, the local radius $h(x,t)$ of the fluid neck goes
to zero; this point is a singularity of the underlying equation of motion.
Since the drop breaks into two pieces, there is no way the problem can be 
continued without generalising the formulation to one that includes 
topological changes. However, in this review we adopt a broader view of what
constitutes a singularity. We consider it as such whenever there is a 
loss of regularity, which implies that there is a length scale which goes
to zero. This is the situation under which one expects self-similar 
behaviour, which is our guiding principle. 
\begin{figure}[t]
\centering
\includegraphics[width=0.2\hsize]{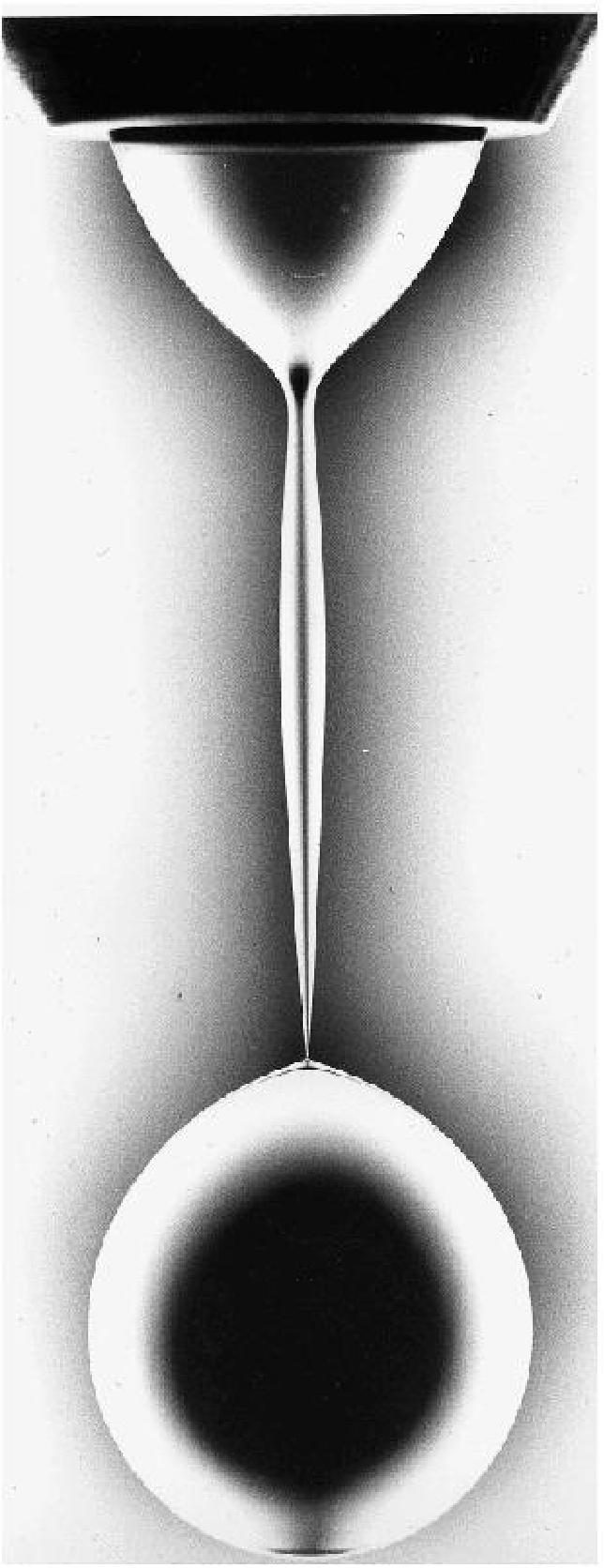}
\caption{A drop of Glycerin dripping through Polydimethylsiloxane 
near pinch-off \protect\cite{CBEN99}. The nozzle diameter is $0.48$ cm, the
viscosity ratio is $\protect\lambda=0.95$. }
\label{lava}
\end{figure}

A fascinating aspect of the study of singularities is that they describe a
great variety of phenomena which appear in the natural sciences and beyond
\cite{K97}. Some examples of such singular events occur in free-surface flows
\cite{E97}, turbulence and Euler dynamics (singularities of vortex tubes
\cite{M00,GMG98} and sheets \cite{CFMR05}), elasticity \cite{AB03},
Bose-Einstein condensates \cite{BR02}, non-linear wave physics \cite{MGF03},
bacterial growth \cite{HV96a,BCKSV99}, black-hole cosmology \cite{C93,MG03},
and financial markets \cite{Sor03}.

In this paper we consider evolution equations
\begin{equation}
h_t=F[h],  \label{ge}
\end{equation}
where $F[h]$ represents some (nonlinear) differential or integral operator.
We will also discuss cases where $h$ is a vector, and thus (\ref{ge}) is a
system of equations. Furthermore, the spatial variable $x$ may also have 
several dimensions, and thus potentially different scaling in different
coordinate directions. We will cite some examples below, but few of 
the higher-dimensional cases have so far been analysed in detail. 
For the purpose of the following discussion, let us suppose that both
$x$ and $h$ are scalar quantities, and that the singularity occurs at  
a single point in space and time $x_0,t_0$. If $t^{\prime }= t_0-t$ and 
$x^{\prime }= x-x_0$, we are looking for local solutions of (\ref{ge}) which
have the structure
\begin{equation}
h(x,t)= t^{\prime \alpha} H(x^{\prime }/t^{\prime \beta}),  
\label{ss}
\end{equation}
with appropriately chosen values of the exponents $\alpha,\beta$.
Note that later the prime is also used to indicate a derivative. 
However, this will always be with respect to a spatial variable 
like $x,z$, or the similarity variable $\xi$, hence confusion 
should not arise. 

Giga and Kohn \cite{GK85,GK87} proposed to introduce self-similar variables $%
\tau=-\ln(t^{\prime })$ and $\xi=x^{\prime }/t^{\prime \beta}$ to study the
asymptotics of blow up. Namely, putting
\begin{equation}
h(x,t)=t^{\prime \alpha}H(\xi,\tau),  \label{simg}
\end{equation}
(\ref{ge}) is turned into the ``dynamical system''
\begin{equation}
H_{\tau}=G[H]\equiv \alpha H - \beta\xi H_{\xi} + F[H].
\label{ds}
\end{equation}
By virtue of (\ref{ds}), solutions to the original PDE (\ref{ge}) 
for given initial data can be viewed as orbits in some infinite 
dimensional phase phase, for instance, $L^2$. To understand the 
blow-up of (\ref{ge}), Giga and Kohn proposed to study the 
long-time behaviour of the {\it dynamical system} (\ref{ds}). 
Thus in particular, one is interested in the attractors of (\ref{ds})
($\omega$-limit sets in the notation which is customary in the 
context of partial differential equations, see \cite{GV04} and 
references therein). If (\ref{ss}) is indeed a solution of (\ref{ge}), 
the right hand side of (\ref{ds}) is \textit{independent} of $\tau$, 
and self-similar solutions of the form (\ref{ss}) are 
\textit{fixed points} of (\ref{ds}), which we will denote by 
$\overline{H}(\xi)$. By studying the dynamics close 
to the fixed point, we find that the dynamical system (\ref{ds}) 
frequently reduces to very few dimensions. Thus on one hand
one obtains detailed information on the behaviour of the original 
problem (\ref{ge}) near blowup. On the other hand, one also gains a 
fruitful means of \textit{classifying},
or at least \textit{characterising} singularities. 

The most basic linear stability analysis of this self-similar
solution consists in linearising around the fixed point according to 
\begin{equation}
H = \overline{H}(\xi) + \epsilon P(\xi,\tau), 
\label{ds_lin}
\end{equation}
which gives 
\begin{equation}
P_{\tau}={\cal L} P ,
\label{sd_lin}
\end{equation}
where ${\cal L} \equiv {\cal L}(\overline{H})$ depends on the 
fixed point solution $\overline{H}$.
To solve (\ref{sd_lin}), we write $P$ as a superposition of 
eigenfunctions $P_j$ of the operator ${\cal L}$:
\begin{equation}
P(\xi)=\sum_{j=1}^{\infty} a_j(\tau) P_j(\xi),
\label{exp_p}
\end{equation}
where $\nu_j$ is the eigenvalue:
\begin{equation}
{\cal L} P_j = \nu_j P_j.
\label{mean_lyn}
\end{equation}

In the cases we know, the spectrum turns out to be discrete. 
For evolution PDE's involving second order elliptic
differential operators, such as semilinear parabolic equations, 
mean curvature or Ricci flows, the discreteness of the spectrum of 
the linearisation about the fixed point is a direct consequence of 
Sturm-Liouville theory \cite{Coddington,Levitan}. 
This theory establishes that, 
under quite general conditions on the coefficients of a second 
order linear differential operator and the boundary conditions, 
its spectrum is discrete and the corresponding
eigenfunctions form a complete set in a suitably weighed $L^2$ space.
Some explicit examples are presented in subsection \ref{quad}. 
For general linear operators such a theory is not available, and one
has to study the spectrum case by case. 

Now the solution of (\ref{sd_lin}) corresponding to
$P_j$ is 
\begin{equation}
P = e^{\nu_j\tau}P_j,
\label{eigen_exp}
\end{equation}
and all eigenvalues need to be 
negative for the similarity solution to be stable. In that case,
convergence to the fixed point is exponential, or {\it algebraic} 
in the original time variable $t'$. Soon the solution has effectively 
reached the fixed point, and there is very little change in the 
self-similar behaviour. If one or several of the eigenvalues
around the fixed point vanish, the approach to the fixed point is 
slow, and the dynamics is effectively described by a dynamical system
whose dimension corresponds to the number of vanishing eigenvalues. 
The same holds true if the attractor has few 
dimensions (such as a limit cycle or a low-dimensional chaotic attractor). 
Thus although singular behaviour is in principle a problem to be solved 
in infinite dimensions, in practise it typically reduces to a dynamical 
problem of few dimensions. In this review we analyse singularities from the
point of view of the {\it slow dynamics} contained in (\ref{ds}),
to obtain an overview and tentative classification of possible 
scaling behaviours. We also emphasise the physical significance of 
these different types of behaviours. 

The perspective described above suggests a close relationship to
the description of scaling phenomena by means of the renormalisation 
group, developed in the context of critical phenomena 
\cite{Goldenfeld93,BKL94}; we will continue to point out similarities, 
but we are not aware that a classification similar to ours has been 
achieved using the language of the renormalisation group. For a 
computational perspective on analysing (\ref{ds}) in terms of its
slow dynamics, see \cite{CDGK04}. Finally, another approach sometimes 
associated with the classification of singularities is catastrophe 
theory \cite{Arnold84}. 
However, as far as we are aware catastrophe theory only yields useful
results if the problem can be mapped onto a low-dimensional geometrical
problem, which can in turn be rephrased in terms of normal forms of
polynomials. This has been shown to be the case for wave problems 
such as shock formation and wave breaking \cite{PLGG08}, as well as
singularities of the eikonal equation \cite{AA03} and related 
problems \cite{Berry07}. 

In this paper we discuss the following cases:
\begin{enumerate}
\item[(I)] \textit{Stable fixed points} (section \ref{fixed}) 

In this case the fixed point is approached exponentially in 
the logarithmic variable $\tau$, so the dynamics is described
by the self-similar law (\ref{ss}). This pure power-law
behaviour is also known as type-I self-similarity \cite{AV97}.

\item[(II)] \textit{Centre manifold} (section \ref{sec:centre})

Here one or more of the eigenvalues around the fixed point are zero.
As a result, the approach to the fixed point is only algebraic, leading
to logarithmic corrections to scaling. This is called type-II self-similarity 
\cite{AV97}; it characterises cases where the blow-up rate is different 
from what is expected on the basis of a solution of the type (\ref{ss}). 

\item[(III)] \textit{Travelling waves} (section \ref{travel})

Solutions of (\ref{ge}) converge to $h=t^{\prime \alpha}\phi(\xi+c\tau)$,
which is a travelling wave solution of (\ref{ds}) with propagation velocity 
$c$.

\item[(IV)] \textit{Limit cycles} (section \ref{limit})

Solutions have the form $h={t^{\prime}}^{\alpha }\mathbf{\psi }%
\left[\xi,\tau\right]$ with $\psi$ being a periodic function of period $T$
in $\tau$. This is known as ``discrete self-similarity'' \cite{C93,MG07},
since at times $\tau_{n}=\tau_{0}+nT$, n integer, the solution looks 
like a self-similar one.

\item[(V)] \textit{Strange attractors} (section \ref{sec:strange})

The dynamics on scale $\tau$ are described by a nonlinear 
(low-dimensional) dynamical system, such as the Lorenz equation. 

\item[(VI)] \textit{Multiple singularities} (section \ref{multiple})

Blow-up may occur at several points $(x_0,t_0)$ (or indeed in any set
of positive measure), in which case the description (\ref{ds}) is not 
useful. We also describe cases where (\ref{ss}) still applies, 
and blow-up occurs at a single point, but the underlying dynamics
is really one of two singularities which merge at the singular time.

\end{enumerate}

\begin{table}
  \begin{center}
    \leavevmode
\begin{tabular}{|l|l|l|l|}
\hline
Equation & Type & Dynamics & Section \\ \hline\hline
\multicolumn{4}{|c|}{Free surface flow} \\ \hline\hline
$h_t + \nabla\cdot(h^n \nabla\triangle h) \pm \nabla(h^p\nabla h)=0$ & I,II &
stable ? & \ref{thin} \\ \hline
$(h^2)_t + (h^2u)_x =0$ & I &  &  \\
$\rho(u_t + uu_x) = (h^2u_x)_x/h^2-(h^{-1})_x$ &  
 & stable & \ref{thin} \\ \hline
$h_t=\left[h\kappa_x/(1+h_x^2)^{1/2}\right]_x,$ & I & 
&  \\
$\kappa = 1/(h(1+h_x^2)^{1/2}) - h_{xx}/(1+h_x^2)^{3/2}$ &  &
stable & \ref{sub:first} \\ \hline
$h_t + (hu)_x =0, \quad u_t + uu_x = h_{xxx}$ & 
    I & stable & \ref{more} \\ \hline
$\int\frac{\ddot{a}(\xi,t)d\xi}{\sqrt{(x-\xi)^2+a(x,t)}} = \frac{\dot{a}^2}{%
2a} $ & II & $v_{\tau}=-v^3$ & \ref{cavity} \\ \hline
$u(x)=\frac{1}{4} \int\frac{h_z(z)}
{\sqrt{h^2(z)+(x-z)^2}}dz$ &  &  &  \\
$(h^2)_t + (h^2u)_x = 0 $ & III & stable & \ref{travel} \\ \hline\hline
\multicolumn{4}{|c|}{Geometric evolution equations} \\ \hline\hline
$h_t= h_{zz}/(1+h_z^2) - 1/h$ & II & $u_{\tau}=-u^2$ & \ref{quad} \\ \hline
$\psi_t = \psi_{ss} - 
(n-1)(1-\psi_s^2)/\psi$ & II & $u_{\tau}=-u^2$ & \ref{quad} \\ \hline\hline
\multicolumn{4}{|c|}{Reaction-diffusion equations} \\ \hline\hline
$u_t - \triangle u = f(u)$ & II & $u_{\tau}=-u^2$ & \ref{sub:react} \\ \hline
$u_t - \nabla\cdot(|u|^m\nabla u) = u^p$ & II & unknown
 & \ref{sub:react} \\ \hline
$\rho_t+\nabla\cdot(\rho\nabla S-\nabla\rho) = 0, 
\quad \rho=-\triangle S$ & II &
$u_{\tau}=-u^3$ & \ref{sub:KS} \\ \hline\hline
\multicolumn{4}{|c|}{Nonlinear dispersive equations} \\ \hline\hline
$u_t + uu_x = 0$ & I & stable & \ref{shock} \\ \hline
$i\psi_t + \triangle\psi + |\psi|^p\psi=0$ & I,II & 
$u_{\tau}=-u^2/v$ & \\ &  & $v_{\tau}=-uv$ & \ref{sub:NLSE} \\ \hline
$u_t+u^p u_x+u_{xxx}=0$ & II & unknown & \ref{other_disp} \\ \hline
$u_t-u_{xxt} + 3uu_x = 2 u_xu_{xx} + uu_{xxx}$ & I 
& unknown & \ref{other_disp} \\ \hline
$u_t = 2fv,\quad v_t=-2fu,\quad f_t=f^2 $ & IV 
 & circle & \ref{limit}  \\ \hline
Choptuik equations & I, IV & limit cycle & \ref{limit} \\ \hline\hline
$u_{tt}=u_{xx} + |u|^p u$ & I,II & unknown & \ref{sub:semi} \\ \hline
\multicolumn{4}{|c|}{Fluid equations} \\ \hline\hline
$u_t + (u\cdot\nabla)u = -\nabla p + \triangle u, 
\quad \nabla\cdot u = 0$ 
& I, IV ? & unknown & \ref{sub:NS} \\ \hline
$u_t + (u\cdot\nabla)u = -\nabla p, \quad \nabla\cdot u = 0$ & I, IV ? 
& unknown & \ref{sub:NS} \\
\hline
$u_t + u u_x + vu_y = -p_x + u_{yy}, \quad u_x+v_y=0$ 
& I & stable & \ref{sub:NS} \\
\hline
\end{tabular}
\end{center}
\caption{A summary of PDE's discussed in this paper.
The first column gives the PDE in question, the second the type of
dynamics near the fixed point according to the classification
enumerated above. In the case of attracting fixed-point dynamics,
it is classed as ``stable'', otherwise the equation governing
the slow dynamics is given.}
\label{list}
\end{table}

This paper's aim is to assemble the body of knowledge on singularities of
equations of the type (\ref{ge}) that is available in both the mathematical
and the applied community, and to categorise it according to the types given
above. In addition to rigorous results we pay particular attention to
various phenomenological aspects of singularities which are often crucial
for their appearance in an experiment or a numerical simulation. For
example, what are the observable implications of the convergence onto
the self-similar form (\ref{ss}) being slow? In most
cases, we rely on known examples from the literature, but the problem
is almost always reformulated to conform with the formulation advocated 
above. However, some examples are entirely new, which we will indicate 
as appropriate. For each of the above categories, we will present at
least one example in greater detail, so the analysis can be followed
explicitely. A concise overview of the equations presented in this 
review is given in Table~\ref{list}. 

\section{Stable fixed points}
\label{fixed}
A sub-classification into self-similarity of the {\it first and 
second kind} has been expounded in \cite{BZ72,Sedov,Sachdev,B96}.
Self-similar solutions are of the first kind if (\ref{ss}) only 
solves (\ref{ge}) for one set of exponents $\alpha,\beta$; their 
values are fixed by either dimensional analysis or symmetry, and are 
thus rational. Solutions are of the second kind if solutions (\ref{ss}) exist
{\it locally} for a continuous set of exponents $\alpha,\beta$; however,
in general these solutions are inconsistent with the boundary or initial 
conditions. Imposing these conditions leads to a non-linear eigenvalue 
problem, whose solution yields irrational exponents in general. 

\subsection{Self-similarity of the first kind}
\label{sub:first}
\begin{figure}[t]
\centering
\includegraphics[width=0.4\hsize]{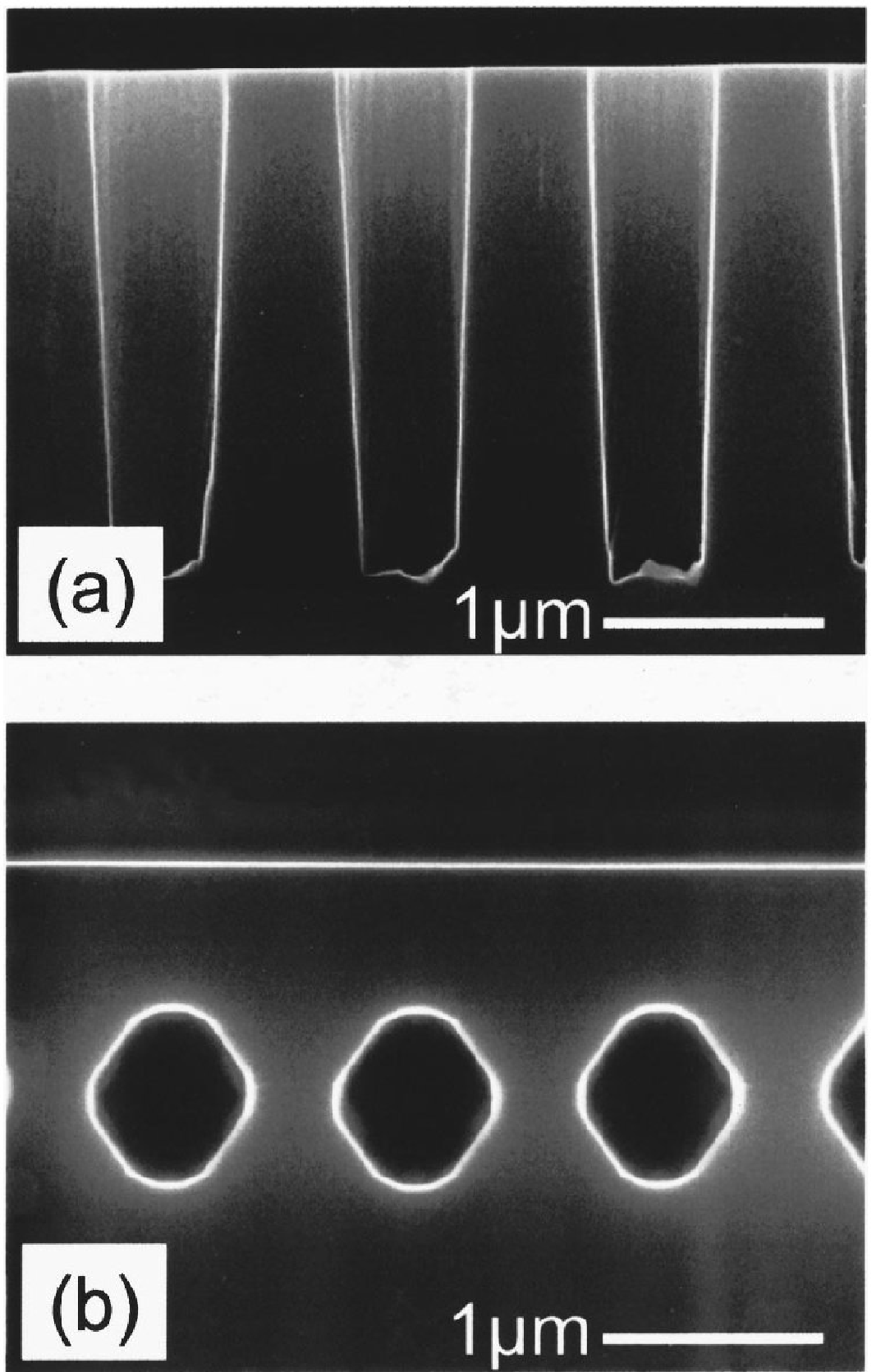}
\caption{\label{trough} 
SEM images illustrating the pinch-off of a row of rectangular 
troughs in silicon (top) \cite{MSTT00}. The bottom picture 
shows the same sample after 10 minutes of annealing at $1100^{\circ}$C.
The troughs have pinched off to form a row of almost spherical 
voids. The dynamics is driven by surface diffusion. 
   }
\end{figure}

Our example, exhibiting self-similarity of the first kind \cite{B96}, 
is that of a solid surface 
evolving under the action of surface diffusion. Namely, 
atoms migrate along the surface driven by gradients of 
chemical potential, see Fig.\ref{trough}. The resulting equations in the 
axisymmetric case, where the free surface is described
by the local neck radius $h(x,t)$, are \cite{NM65}:
\begin{equation}
h_t=\frac{1}{h}\left[\frac{h}{(1+h_x^2)^{1/2}}\kappa_x\right]_x,
\label{sd}
\end{equation} 
where 
\begin{equation}
\kappa = \frac{1}{h(1+h_x^2)^{1/2}} - \frac{h_{xx}}{(1+h_x^2)^{3/2}} 
\label{mean}
\end{equation} 
is the mean curvature. In (\ref{sd}),(\ref{mean}), all lengths 
have been made dimensionless using an outer length scale $R$ (such
as the initial neck radius), and the time scale $R^4/D_4$, where
$D_4$ is a forth-order diffusion constant. 

Physically, it is important to point out that (\ref{sd}) describes 
the evolution of the free surface at elevated temperatures, above
the so-called roughening transition. This implies that the solid
surface is smooth and does not exhibit facets, coming from the 
underlying crystal structure. Above the roughening transition, a
continuum description is still possible \cite{Spohn93}. The study
of these models has lead to a number of interesting similarity solutions
describing singular behaviour of the surface, such as grooves \cite{SAM05}
or mounds \cite{MAS04,MFAS06}. 

\begin{figure}
\centering
\includegraphics[width=0.6\hsize]{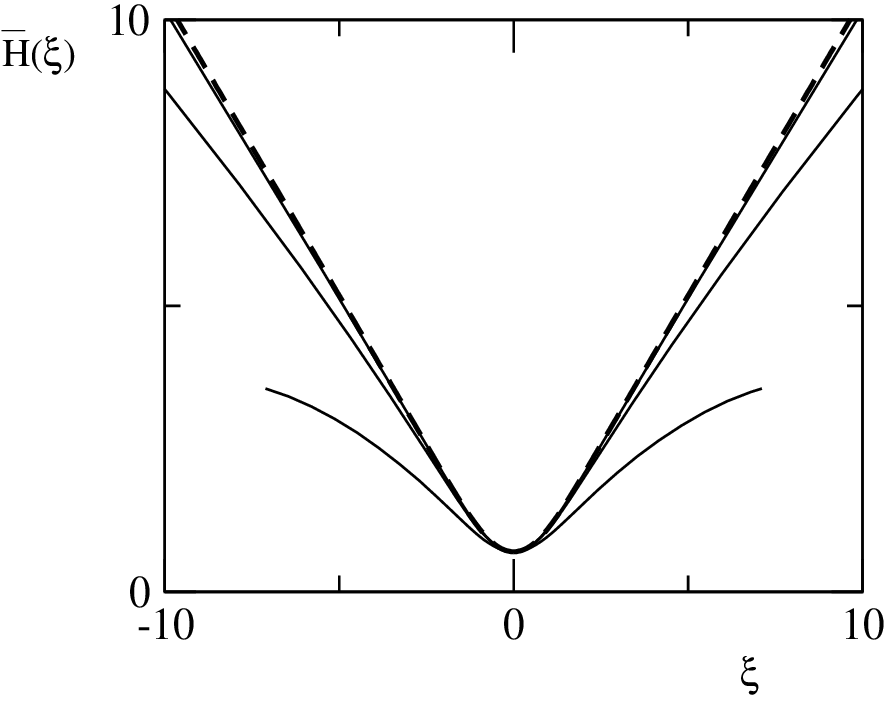}
\caption{\label{sdfig} 
The approach to the self-similar profile for equation (\ref{sd}).
The dashed line is the stable similarity solution $H(\xi)$
as found from (\ref{sdsim}). The full lines are rescaled profiles 
found from the original dynamics (\ref{sd}) at 
$h_m = 10^{-1},10^{-2}$, and $h_m=10^{-3}$, respectively.
As the singularity is approached, they converge rapidly onto the 
similarity solution (\ref{ssd}).
   }
\end{figure}

At a time $t' \ll 1$ away from breakup, dimensional analysis
implies that $\ell=t'^{1/4}$ is a local length scale. 
This suggests the similarity form 
\begin{equation}
h(x,t)= t'^{1/4}H(x'/t'^{1/4}),
\label{ssd}
\end{equation}
and thus the exponents $\alpha,\beta$ of (\ref{ss}) are fixed by
dimensional analysis, which is typical for self-similarity of the first kind.
Of course, the result (\ref{ssd}) also follows when directly searching for a 
solution of (\ref{sd}) in the form of (\ref{ss}). 
In other cases, a unique set of local scaling exponents is determined
by symmetry \cite{Edrop05}. The similarity form
of the PDE becomes 
\begin{equation}
-\frac{1}{4}(H-\xi H_{\xi}) = \frac{1}{H}
\left[\frac{H}{(1+H_{\xi}^2)^{1/2}}
\kappa_{\xi}\right]_{\xi}, \quad \xi=\frac{x'}{t'^{1/4}}
\label{sdsim}
\end{equation}
where $\kappa$ is the mean curvature of $H$. 

Solutions of (\ref{sdsim}) have been studied extensively in \cite{BBW98}. 
To ensure matching to a time-independent outer solution, the leading order 
time dependence must drop out from (\ref{ssd}), implying that 
\begin{equation}
H(\xi) \sim c|\xi| , \quad \xi \rightarrow \pm\infty;
\label{bound}
\end{equation}
the general form of this matching condition for self-similar 
solutions of the form (\ref{ss}) is 
\begin{equation}
H(\xi) \sim c|\xi|^{\frac{\alpha}{\beta}} , 
\quad \xi \rightarrow \pm\infty. 
\label{bound_gen}
\end{equation}
All solutions of the similarity equation (\ref{sd}), and which obey
the growth condition (\ref{bound}) are symmetric, and form a  
discretely infinite set \cite{BBW98}, similar to a number of other
problems discussed below. The series of similarity solutions is 
conveniently ordered by descending values of the minimum,
see table \ref{series}. Only the lowest order solution $H_0(\xi)$
is stable, and is shown in Fig.~\ref{sdfig}; we return to the issue of 
stability in section \ref{stable} below. The fact that permissible
similarity solutions form a discrete set implies a great deal of 
``universality'' in the way pinching can occur. It means that the 
local solution is independent of the outer solution, and rather that the
former imposes constraints on the latter; in particular, the prefactor 
$c$ in (\ref{bound}) must be determined as part of the 
solution (see Table~\ref{series}).

\begin{table}
  \begin{center}
    \leavevmode
\begin{tabular}{cccccccccccccccccc}
i & $H_i(0)$ & $c_i$ & \\ \hline
0 & 0.701595 & 1.03714 \\
1 & 0.636461 & 0.29866 \\
2 & 0.456842 & 0.18384 \\
3 & 0.404477 & 0.13489 \\
4 & 0.355884 & 0.10730 \\
5 & 0.326889 & 0.08942 \\
\end{tabular}
\end{center}
\caption{A series of similarity solutions of (\ref{sdsim}) as given
in \cite{BBW98}. The higher-order solutions become successively thinner
and flatter. }
\label{series}
\end{table}

\subsubsection{Thin films and thin jets}
\label{thin}
A further class of solutions displaying self-similarity of the first kind is
the generalised long-wave thin-film equation
\begin{equation}
h_{t}+\nabla \cdot (h^{n}\nabla \Delta h - B h^{m}\nabla h) = 0\ ,\ n>0 .
\label{thin_film}
\end{equation}
The mass flux in this equation has two contributions: 
the first is due to surface tension, and the second is due to an external 
potential. When $n=m=3$, then $z=h(\mathbf{x},t)$ represents the height 
of a film or a drop of viscous fluid over a flat surface, located at 
$z=0$; the external potential is gravity. If $B$ is negative, 
(\ref{thin_film}) describes a film that is hanging from a ceiling. 
Regardless of the sign of $B$, there is no singularity in this case
\cite{Pugh}. The case $n=1$ and $B=0$ 
corresponds to flow between two solid plates, to which we return in 
section \ref{HS_sub} below. 

Solutions to (\ref{thin_film}) are said to develop point singularities if 
$h$ goes to zero in finite time. This happens if one incorporates
van der Waals forces, which at leading order implies $n=3$ and $m=-1$
with $B<0$. 
In \cite{ZL99b}, \cite{WB00} (see also the review \cite{BG05}, where
further full numerical simulations and mathematical theory are reported) 
the existence of radially symmetric self-similar touchdown solutions 
of the form
\begin{equation}
h(r,t)=t'^{\frac{1}{5}}H(\xi), \quad \xi= r/t'^{\frac{2}{5}}
\label{touch}
\end{equation}
is shown numerically in this case. Self-similar solutions that 
touch down along a 
line exist as well, but they are unstable. A proof of formation of 
singularities in this context has been provided by Chou and Kwong \cite{CK07}.

A related set of equations are those for thin films and jets, 
but which are isolated instead of being in contact with a solid. 
Problems of this sort furnish many examples of type-I scaling, 
as reviewed from a physical perspective in \cite{EV08}. 
If the motion is no longer dampened by the presence of a solid, 
inertia often has to be taken into account. This means that a 
separate equation for the velocity is needed, which is essentially 
the Navier-Stokes equation below, but often simplified by a reduction
to a single dimension. 
Thus one has solutions of the form 
\begin{equation} \label{gen_sol}
h(x,t) = t'^{\alpha}H(\xi), \quad 
u(x,t) = t'^{\beta-1}U(\xi), 
\end{equation}
where $\xi = x'/t'^{\beta}$. If $\alpha > \beta$ the profile is 
slender, and the dynamics is well described in a shallow-water
theory. In this case the equations for an axisymmetric jet with
surface tension become
\begin{equation}\label{hequ}
\partial_{t}h^2 + \partial_x(uh^2) = 0
\end{equation}
and 
\begin{equation}\label{vequ}
\rho(\partial_t u + u\partial_x u) = 
-(\gamma/\rho)\partial_x(1/h) + 3\nu\frac{\partial_x(\partial_xuh^2)}{h^2} . 
\end{equation}

The system (\ref{hequ}),(\ref{vequ}) is interesting because it exhibits
different scaling behaviours depending on the balance between the
three different terms in (\ref{vequ}) \cite{Edrop05}. This is an
illustration of the principle of {\it dominant balance}, which is of
great practical importance in practise, where it is a priori not 
known which physical effect will be dominant. In the case of (\ref{vequ}),
these are the forces of inertia on the left, surface tension
(first term on the right), and viscosity (second term on the right).
Pinching is driven by surface tension, so it must always be part
of the balance. Three different possible balances remain \cite{Edrop05}:

(i)
In the first case \cite{E93}, all forces in (\ref{vequ}) are 
balanced as the singularity is approached. The exponents 
$\alpha=1,\beta=1/2$ in (\ref{gen_sol}) follow directly from
this condition. As shown in \cite{BLS96},
there is a discretely infinite sequence of self-similar 
profiles $H(\xi),U(\xi)$ corresponding to this balance. Numerical 
evidence strongly suggests that only the first profile, corresponding
to the thickest thread, is stable \cite{E97}. All the other profiles
are unstable, and thus cannot be observed. We will revisit this
general scenario again below, when we study the stability of
fixed points more generally. 

(ii)
The second possibility corresponds to a balance between surface tension and
viscous forces, thus putting $\rho=0$ in (\ref{vequ}). Physically,
this occurs if the fluid is very viscous \cite{P95}. In section 
\ref{viscous} below we will describe the pinching solution 
corresponding to this case in more detail, as an example of 
self-similarity of the second kind. The exponent $\alpha=1$ is
fixed by the balance, but $\beta$ is fixed only by an integrability 
condition. This once more results in an infinite sequence of 
solutions, ordered by the value of $\beta$. Again, only one profile, 
which has the largest value of $\beta = 0.17487$ is stable. This time,
this corresponds to the smallest value of the minimum radius $R_0$, 
or the thinnest thread, as opposed to thickest thread in the case
of the inertial-surface tension-viscous balance. 

If one inserts this viscous solution into the original
equation (\ref{vequ}), one finds that in the limit $t'\rightarrow 0$,
the inertial term on the left grows faster than the two terms 
on the right. This means that regardless how large the viscosity,
eventually all three terms become of the same order, and one
observes a crossover to the inertial-surface tension-viscous 
similarity solution described above, which is characterised 
by another set of scaling exponents and similarity profiles.
In particular, the surface tension-viscous solution is symmetric
about the pinch point, whereas the solution containing inertia
is highly asymmetric \cite{E05}. We remark that crossover 
between different similarity solutions may also occur by another
mechanism, not directly related to the dominant balance between
different terms in the equation (cf. section \ref{HS_sub}).

Equations (\ref{hequ}),(\ref{vequ})
correspond to a viscous liquid, surrounded by a gas, which is 
not dynamically active. The case of an external viscous fluid 
is considered in detail in section \ref{travel} below. The 
case of no internal fluid is special, in that the dynamics 
decouples completely into one for independent slices \cite{DCZSHBN03}. 
As a result, there is no universal profile associated with the 
breakup of a bubble in a viscous environment, but rather it is 
determined by the initial conditions. 

(iii)
At very low viscosity ($\nu \approx 0$ in (\ref{vequ})), the relevant 
balance is one where inertia is balanced by surface tension, so
one might want to set $\nu=0$ in (\ref{vequ}), as done originally
in \cite{TK90}. However, the resulting equations do not lead
to a selection of the values of the scaling exponents $\alpha,\beta$;
instead, there is a continuum of solutions \cite{FV99}, parameterised by the
value of $\alpha$, each with a continuum of possible similarity
profiles. In fact, for vanishing viscosity (\ref{hequ}),(\ref{vequ})
does not go toward a pinching solution, but the slope of the interface
steepens, and one finds a shock solution \cite{E00}, similar to
the generic scenario described in section \ref{shock} below.

It was however shown numerically in \cite{CS97,DHL98}, and investigated in 
more detail in \cite{LL03}, that pinch-off of an inviscid fluid
is well described by a solution of the full three-dimensional, 
axisymmetric potential flow equations. This is thus an example of a 
similarity solution of higher order in the independent variable, but
both coordinate directions scale in the same way. The scaling exponents in 
(\ref{gen_sol}) are $\alpha = \beta=2/3$ in this case, which violates
the assumption $\alpha > \beta$ for the validity of the  
shallow water equations (\ref{hequ}),(\ref{vequ}). In addition,
we note that the similarity profile can no longer even be written 
as a graph as assumed in (\ref{gen_sol}), but {\it turn over}, as 
first observed experimentally in \cite{L87}.
It is not known whether there also exists a {\it sequence}
of similarity solutions, as in the case of the other balances. 
The case of no internal fluid is again very special, and leads
to type-II scaling. It is considered in section \ref{cavity} below. 

Finally, variations of (\ref{hequ}),(\ref{vequ}) have been investigated
in \cite{VLW01a}. Breakup was considered in arbitrary dimensions $d$
(yet retaining axisymmetry) and with the pressure term $1/h$ replaced
by an arbitrary power law $1/h^p$. After introducing a new variable
$1/h^p$, there remains a single parameter $r = (d-1)/p$, which can 
formally be varied continuously. For all values of $r$, discrete 
sequences of type-I solutions are obtained. For $r >1/2 $, profiles
are asymmetric, while below that value they are symmetric. At the 
critical value, both types of solutions coexist. Another interesting
feature of the limit $r = 1/2$ is that the viscous term becomes
subdominant at leading order. However, similar to the case $d=3,p=1$
mentioned above, no selection takes place in the absence of the 
viscous term. Nevertheless, the solutions selected by the presence 
of the viscous term are very close to an appropriately chosen member
of the family of inviscid solutions. 

\subsection{Singularities in Euler and Navier-Stokes equations}
\label{sub:NS}
One of the most important open problems, both in physics and mathematics, is
the existence of singularities in the equations of fluid mechanics: Euler
and Navier-Stokes equations in three space dimensions. The Navier-Stokes 
equations represent the evolution of a viscous incompressible fluid and 
are of the form
\begin{equation}
\mathbf{u}_{t}+\mathbf{u}\cdot \nabla \mathbf{u} =-\nabla p+Re^{-1} \Delta
\mathbf{u}, \qquad \nabla \cdot \mathbf{u} =0,
\label{NS}
\end{equation}
where $\mathbf{u}$ represents the velocity field, $p$ the pressure in the
fluid and Re is a dimensionless parameter called Reynolds number. Formally,
by making Re$\rightarrow \infty $, the term involving $\Delta \mathbf{u}$
vanishes and we arrive at the Euler system, that models the evolution of the
velocity and pressure fields of an inviscid incompressible fluid:
\begin{equation}
\mathbf{u}_{t}+\mathbf{u}\cdot \nabla \mathbf{u} =-\nabla p,\qquad
\nabla \cdot \mathbf{u} =0.
\label{EU}
\end{equation}
We exclude from our discussion certain ``exact'' blow-up solutions 
of the Euler equations \cite{GMS03}, which have the defect
that the velocity goes to infinity {\it uniformly} in space; in
other words, they lack the crucial mechanism of {\it focusing}.
Formally, they are of course similarity solutions of (\ref{EU}), 
but with spatial exponent $\alpha=0$. 

As we mentioned above, the existence of singular solutions is unknown.
Nevertheless, some scenarios have been excluded. For the 
Navier-Stokes equations, there exists no nontrivial self-similar solution 
of the first kind
\begin{equation}
\mathbf{u}(\mathbf{x},t)=t'^{-1/2}\mathbf{U}\left(
\mathbf{\xi}\right), \quad \mathbf{\xi} = {\bf x}'/t'^{1/2}
\label{Leray}
\end{equation}
in $L^{2}(\mathbb{R}^{3})$. This was proved by Necas, Ruzicka and Sverak
\cite{NPS96}. However, this does not exclude the formation of a singularity in 
a localised region: the matching condition (\ref{bound_gen}) for this 
case implies $|{\bf U}|\propto|\xi|^{-1}$ as $|\xi|\rightarrow \infty$,
which is {\it not} in $L^2$. Therefore, the theorem \cite{NPS96} does
not apply.

A possible
self-similar solution consisting of two skewed vortex-pairs has been
proposed by Moffatt in \cite{M00} in the spirit of the scenario suggested
by the numerical simulations of Pelz \cite{P97}, of the implosion
of six vortex pairs in a configuration with cubic symmetry. More recent
numerical experiment by Hou and Li \cite{HL08} seem to indicate that,
although the velocity field may grow to very large values, singularities
in the above mentioned scenarios saturate eventually and the solutions
remain smooth. It has been argued in \cite{PS05} that no self-similar
solutions for Euler system should exist and that the "limit-cycle" scenario
described in section \ref{limit} could apply.

Under certain circumstances, such as special symmetry conditions or
appropriate asymptotic limits, the Navier-Stokes and Euler systems may
simplify and give rise to models for which the question of existence 
of singular solutions is somewhat simpler to analyse. This is the case 
for the Prandtl boundary-layer equations for the 2-D evolution of the 
velocity field $(u,v)$ in $y\geq 0$:
\begin{equation}
u_{t}+uu_{x}+vu_{y}=-p_{x}+u_{yy},\quad u_{x}+v_{y}=0
\label{Prandtl}
\end{equation}
with boundary conditions $u=v=0$; $p$ is a given pressure field and the
behaviour of the velocity field at infinity is prescribed. 
Equation (\ref{Prandtl}) describes the asymptotic limit of the Navier-Stokes
equation near a solid body in the limit of large Reynolds numbers $Re$. 
The variable $x$ measures the arclength along the body, and $Re^{1/2}y$
is the distance from the body. Historically, a lot of attention was 
focused on the {\it stationary} version of (\ref{Prandtl}), considering 
it as an evolution equation in $x$. At some position $x_s$ along the 
body, the so-called Goldstein singularity $v \propto (x_s - x)^{-1/2}$
is encountered \cite{G48}, which signals separation of the flow 
from the body. However, in reality the outer flow changes as a result 
of the appearance of a stagnation point, and one has to consider the 
{\it interaction} between the boundary layer and the outer flow \cite{Sychev}.

It is thus conceptually simpler to consider the case of {\it unsteady} 
boundary layer separation, which is described by the first singularity
of (\ref{Prandtl}) at time $t_0$. The formation of singularities of 
(\ref{Prandtl}) in finite time was proved by E and Engquist \cite{EE00}.
It was first found numerically by van Dommelen and Shen \cite{VS80}, and
its analytical structure was investigated in \cite{VS82}, using Lagrangian
variables, which follow fluid particles as they separate from the surface
(see also \cite{CS00}). In the original Eulerian variables, the
self-similar structure is \cite{ECS83,CSW96}
\begin{equation}
\label{u_sep}
u = -u_0 + t'^{1/2}\phi_0^{1/2}U(\xi,\eta), \quad
\xi = \frac{x'-u_0 t'}{t'^{3/2}\phi_0^{1/2}}, \;
\eta = \frac{y\phi_0^{1/4}}{t'^{1/4}\Lambda},
\end{equation}
where $u_0, \phi_0$, and $\Lambda$ are constants which depend
on the problem, while $U$ is universal and can be given in terms
of elliptic integrals. Note that the exponents for $u$ and $x$
are the generic exponents for a developing shock (see section \ref{shock} 
below), while the similarity exponent in the $y$-direction is different 
from the scaling for two-dimensional breaking waves \cite{PLGG08}. 
We stress that the appearance of a singularity in (\ref{Prandtl})
does not mean that the full 2D Navier-Stokes equation has developed a 
singularity. Instead, lower order terms in the asymptotic expansion that
lead to (\ref{Prandtl}) become important close to the singularity.

In relation with singularities in fluid mechanics, we can mention briefly a
few important problems involving models or suitable approximations to the
original Euler and Navier-Stokes systems. One concerns weak solutions to
the Euler system for which the vorticity ($\omega =\nabla \times \mathbf{u}$) 
is concentrated in curves or surfaces. This is the case of the so called vortex
filaments and sheets in which the vorticity remains concentrated for all
times, in absence of viscosity. 
A useful way to represent the vortex sheet, when it evolves in 2D, is
by assuming the location of its points $(x(\alpha ,t),y(\alpha ,t))$ as
complex numbers $z(\alpha ,t)=x(\alpha ,t)+iy(\alpha ,t)$. Then, the
evolution of $z(\alpha ,t)$ is given by the so-called 
Birkhoff-Rott equation \cite{Saff_book}:
\begin{equation}
z_{t}^{\ast }(\alpha ,t)=\frac{1}{2\pi i}PV\int_{-\infty }^{\infty }\frac{%
\gamma (z(\alpha ^{\prime },t),t)}{z(\alpha ,t)-z(\alpha ^{\prime },t)}%
z_{\alpha }(\alpha ^{\prime },t)d\alpha ^{\prime }\ ,  \label{kh1}
\end{equation}
where $z^{\ast }$ stands for the complex conjugate of $z$. The 
principal value is denoted by PV, and $\gamma $ is the
vortex strength and is such that $d\Gamma =\gamma (z(\alpha ,t),t)z_{\alpha
}(\alpha ,t)d\alpha $ is constant along particle paths of the flow. The
question then is whether or not these geometrical objects will remain smooth
at all times or develop singularities in finite time. In the case of vortex
sheets, singularities are known to develop in the form of a divergence of the
curvature at some point. These are called Moore's singularities after their
observation and description by D. W. Moore \cite{M79}. A mathematical proof
of existence of these singularities is provided by Caflisch and Orellana in 
\cite{CO89}. These singularities exhibit self-similarity of the first kind
as shown, for instance, in \cite{HFV08}: if one
defines the inclination angle $\theta (s,t)$ in terms of the arclength
parameter $s$ as such that $z_{s}=e^{i\theta }$, then the curvature is given
by $\kappa =\theta _{s}$ and may blow-up in the self-similar form (up to
multiplicative constants):
\begin{equation}
\kappa (s,t')=\frac{1}{t'^{\delta }}g(\eta), \quad \eta = 
s'/t'\ ,\ \ 0<\delta <1,  \label{kauto}
\end{equation}%
where 
\begin{equation}
g(\eta )=\frac{1}{(1+\eta ^{2})^{\frac{\delta}{2}}}
\sin(\delta \arctan\eta )\ .  \label{kauto2}
\end{equation}
Interestingly, numerical simulations and Moore's original observations
suggest that, although singular solutions with any $\delta$ are
possible, that the solution with $\delta =\frac{1}{2}$ is preferred.
Thus the generically observed geometry near the singularity is of the form 
$y=\left\vert x\right\vert ^{\frac{3}{2}}$, including the case of 
3D simulations. This poses an interesting "selection
problem" for the $\frac{3}{2}$ power which has not received a definitive
answer so far. 

Another type of solution of (\ref{kh1}) has the from of a double-branched 
spiral vortex sheet \cite{K89}. The explicit form is 
\begin{equation}
z(\beta,t) =  \left\{\begin{array}{l}
        t'^q\beta^{\nu} \quad \beta > 0 \\
        t'^q|\beta|^{\nu} \quad \beta < 0 \quad\; ,
                 \end{array}\right.
\label{kambe}
\end{equation}
where the two cases correspond to the two branches of the spiral. 
The parameter $\beta$ is related to integration variable $\alpha$ 
of (\ref{kh1}) by $d\beta=z_{\alpha} d\alpha$.
The exponents are of the form $\nu=1/2+ib$ and $q=1/2+i\mu b$, 
corresponding to a vortex of radius $r=t'^{1/2}$ collapsing in finite
time. However, in this case the vortex sheet strength is found
to increase exponentially at infinity \cite{K89}.

Vortex filaments result as the limit of a vortex
tube when the thickness tends to zero. The fluid flow around a vortex
filament is frequently approximated by a truncation of the Biot-Savart
integral for the velocity in terms of the vorticity. This leads to a
geometric evolution equation for the filament (see \cite{MB02}, chapter 7,
and references therein) that can be transformed, via Hasimoto
transformation, into the cubic Nonlinear-Schr\"{o}dinger in 1D. This fact
allowed Gutierrez, Rivas and Vega to construct exact self-similar solutions
for infinite vortex filaments \cite{GRV03}.\textbf{\ }One can also consider
the vorticity concentrated in a region separating two fluids of different
density and in the presence of gravitational forces. This is the case of the
surface water waves system for which the existence of singularities is open 
\cite{CW07}.

A different approach in the study of singularities for Euler and
Navier-Stokes equations in three space dimensions relies on the 
development of models that share some of the essential mathematical 
difficulties of the original systems, but in a lower space dimension. 
This is the case of the surface quasi-geostrophic equation popularised 
by Constantin, Majda and Tabak \cite{CMT94}:
\numparts
\begin{eqnarray}
&& \theta_t + {\bf v}\cdot\nabla \theta = 0,  \\
&& {\bf v} = \nabla^{\bot}\psi, \quad \theta=-(-\triangle)^{1/2}\psi,
\end{eqnarray}
\endnumparts
to be solved in $d=2$. This system of equations describes the convection of an 
active scalar $\theta$, representing the temperature, 
with a velocity field which is an integral operator of the scalar itself.
Nevertheless, the mere existence of singular solutions to this equation 
in the form of blow-up for the gradient of $\theta$ is
still an open problem. One-dimensional analogues of this problem, representing
the convection of a scalar with a velocity field, which is the Hilbert
transform of the scalar itself do have singularities in the form of cusps, as
proved in \cite{CCCF05}, \cite{CCF05}. The structure of such
singularities has been described in \cite{HF08} and they are, in fact, of
the type described in the next section, that is of the second kind.

\subsection{Self-similarity of the second kind} \label{second}
In the example of the previous subsection, the exponents can be determined
by dimensional analysis, or from considerations of symmetry, and 
therefore assume rational values. In many other problems, however,
the scaling behaviour depends on external parameters, set for example 
by the initial conditions. In that case, the scaling exponent 
can assume any value. Often, this value is fixed by a compatability 
condition, resulting in an irrational answer. We will call this situation 
self-similarity of the second kind \cite{BZ72,B96}. Since it is 
relatively rare that results are tractable analytically, we mention two 
simple examples for which this is possible, although they do not come 
from time-dependent problems.

The first example is that of viscous flow near a solid corner of 
opening angle $2\alpha$ \cite{Moffatt63}. For analogues of this
problem in elasticity, see \cite{C13,SK58} as well as the discussion
in \cite{B96}. This flow is described by a 
Stokes' equation, whose solution near the corner is expected to be 
\begin{equation}
\psi = r^{\lambda}f_{\lambda}(\theta). 
\label{corner}
\end{equation}
If one of the boundaries is moving, scaling is of the first kind,
and $\lambda = 2$ (the so-called Taylor scraper \cite{Batch67}). 
However, if the flow is driven by two-dimensional 
stirring at a distance from the corner, $\lambda$ is determined
by the transcendental equation 
\begin{equation}
\sin2(\lambda-1)\alpha = -(\lambda-1)\sin2\alpha.
\label{mf}
\end{equation}
If $2\alpha < 146^{\circ}$, (\ref{mf}) 
admits complex solutions, which correspond to an infinite sequence
of progressively smaller corner eddies. Since $\lambda$ is complex,
The strength of the eddies decreases as one comes closer to the corner. 

The second example consists in calculating the electric field 
between two non-conducting 
spheres, where an external electric field is applied in the direction
of the symmetry plane \cite{SVA97}. In this case the electric potential
between the spheres is proportional to $(\rho/(R h))^{\sqrt{2}-1}$,
where $\rho$ is the radial distance from the symmetry axis, $R$ 
the sphere radius, and $h$ the distance between the spheres. Thus
in accordance with the the general ideas of self-similarity of the 
second kind, the singular behaviour is not controlled by the local 
quantity $\rho/h$, but the ``outer'' parameter $R$ comes into 
play as well. We now explain two analytically tractable 
{\it dynamical} examples of self-similarity of the second kind. 

\subsection{Breaking waves in conservation laws}
\label{shock}
\begin{figure}[t]
\centering
\includegraphics[width=0.6\hsize]{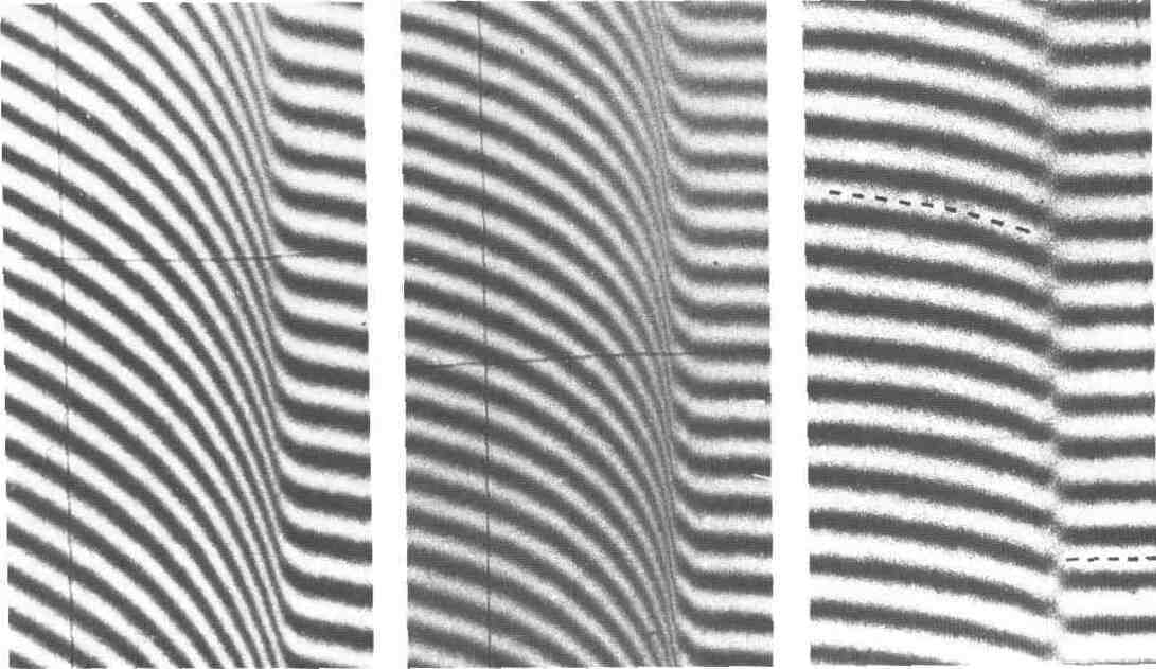}
\caption{Fringe pattern showing the steepening of a wave in a gas, leading
to the formation of a shock, which is travelling from left to 
right \cite{GB54}. The vertical position of a given fringe
is proportional to the density at that point. In the last picture a jump
of seven fringes occurs. 
      }
\label{shock_pic}
\end{figure}
We only consider the simplest model for the formation of a shock wave in gas 
dynamics, which is Burger's equation 
\begin{equation}
u_t + uu_x = 0.
\label{be}
\end{equation} 
It is generally believed that any system of conservation laws
that exhibits blow up will locally behave like (\ref{be}) \cite{Alinhac95}.
For example, Fig.~\ref{shock_pic} shows the steepening of a density 
wave in a gas, leading to a jump of the density in the picture on the 
right. In the words of \cite{GB54}: ``We conclude that an infinite slope in
the theoretical solution corresponds to a shock in real life''. 
As throughout this review, we only consider the dynamics up to the
singularity. Which structure emerges after the singularity depends 
on the regularisation used, as the continuation to times after the 
singularity is not unique \cite{Witham,Dafermos}. 
If the regularisation is diffusive, a 
shock wave forms \cite{Bressan}; if it is a third derivative, 
one finds a KDV soliton.
Finally, regularisation by higher-order nonlinearities has been 
considered in \cite{PLGG08} as a model of wave breaking.

It is well known \cite{LL84a} that (\ref{be}) can be solved exactly
using the method of characteristics. This method consists in 
noting that the velocity remains constant along the characteristic 
curve 
\begin{equation}
z = u_0(x)t + x,
\label{char}
\end{equation} 
where $u_0(x) = u(x,0)$ is the initial condition. Thus 
\begin{equation}
u(z,t) = u_0(x)
\label{char_imp}
\end{equation} 
is an exact solution to (\ref{be}), given implicitly. 

It is geometrically obvious that whenever $u_0(x)$ has a negative
slope, characteristics will cross in finite time and produce a 
discontinuity of the solution. This happens when 
$\partial z/\partial x = 0$, which will occur for the first 
time at the singularity time 
\begin{equation}
t_0 = \min\left\{-\frac{1}{\partial_x u_0(x)}\right\},
\label{first}
\end{equation} 
at a spatial position $x = x_m$. This means a singularity will
first form at 
\begin{equation}
x_0 = x_m - \frac{u_0(x_m)}{\partial_x u_0(x_m)}. 
\label{first_x}
\end{equation} 
Since (\ref{be}) is invariant under any shift in velocity, we can 
assume without loss of generality that $u_0(x_m)=0$, and 
thus that $x_0 = x_m$. This means the velocity is zero at the 
singularity. We now analyse the formation of the singularity
using the local coordinates $x',t'$. In \cite{PLGG08}, this was done
by expanding the initial condition $u_0$ in $x'$, 
and using (\ref{char_imp}), using ideas from 
catastrophe theory \cite{Arnold84}. Here instead we use the similarity ideas 
developed in this paper. 

The local behaviour of (\ref{be}) near $t_0$ can be obtained using the scaling 
\begin{equation}
u(x,t) = t'^{\alpha} U\left(x'/t'^{\alpha+1}\right),
\label{be_scal}
\end{equation}
which solves (\ref{be}). The similarity equation becomes
\begin{equation}
-\alpha U + (1 + \alpha)\xi U_{\xi} + UU_{\xi} = 0, 
\label{be_sim}
\end{equation} 
with implicit solution 
\begin{equation}
\xi = -U - C U^{1+1/\alpha}.
\label{be_impl}
\end{equation} 
The special case $\alpha = 0$ has the solution $U = -\xi$, which 
is inconsistent with the matching condition 
(\ref{bound_gen}), and thus has to be discarded. 

We are thus left with a continuum of possible scaling exponents $\alpha>0$,
as is typical for self-similarity of the second kind. A discretely
infinite sequence of exponents $\alpha_n$ is however selected by
the requirement that (\ref{be_impl}) defines a smooth function 
for all $\xi$. Namely, one must have $1+1/\alpha$ odd, or
\begin{equation}
\alpha_i = \frac{1}{2i+2},\quad i=0,1,2\dots,
\label{be_alpha}
\end{equation} 
and we denote the corresponding similarity profile by $U_{i}$.
The constant $C$ in (\ref{be_impl}) must be positive, but is otherwise 
arbitrary. It is set by the initial conditions, which is another hallmark
of self-similarity of the second kind. However, $C$  can be 
normalised to 1 by rescaling $x$ and $U$. We will see in 
section \ref{stable} that the solution with $\alpha_0$,
\begin{equation}
u(x,t) = t'^{1/2} U_0\left(x'/t'^{3/2}\right),
\label{be_alpha1}
\end{equation} 
is the only stable one, all higher-order solutions are unstable.

It is interesting to look at some possible exceptions to the
form of blow-up given above, suggested by \cite{Alinhac95}:
\begin{equation}
u_t + uu_x = u^{\sigma}.
\label{be_alinhac}
\end{equation} 
This equation is also solved easily using characteristics. 
For $\sigma \le 2$ the blow-up is alway of the form (\ref{be_alpha1}),
for $\sigma > 2$ two different behaviours are possible. For 
small initial data $u_0(x)$, a singularity still forms like
(\ref{be_alpha1}), but in addition $u$ may also go to infinity. 
However, there is a boundary between the two behaviours \cite{Alinhac95},
where the slope blows up at the same time that $u$ goes to infinity.
For this case, one expects all terms in (\ref{be_alinhac}) to be
of the same order, giving
\begin{equation}
u(x,t) = t'^{\frac{1}{1-\sigma}}
U\left(\xi)\right), \quad \xi = x'/t'^{\frac{\sigma-2}{\sigma-1}},
\label{be_spec}
\end{equation} 
with similarity equation
\begin{equation}
\frac{U}{1-\sigma} + \frac{\sigma-2}{\sigma-1}\xi U_{\xi} = 
U^{\sigma} - UU_{\xi}.
\label{spec_sim}
\end{equation} 

\begin{figure}[t]
\centering
\includegraphics[width=0.4\hsize]{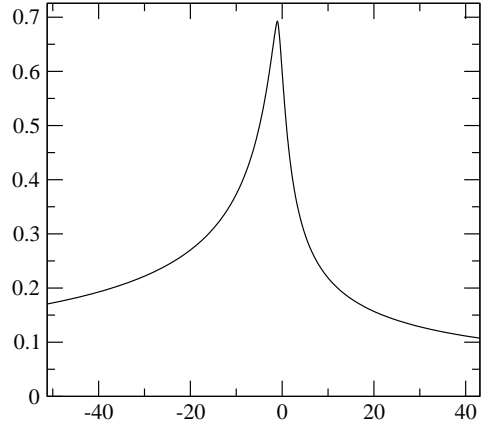}
\caption{The similarity solution (\ref{spec_sol}) for $\sigma=4$.}
\label{alinhac_fig}
\end{figure}
The solution to (\ref{spec_sim}) that has the right decay at 
infinity is 
\begin{equation}
\xi = -\frac{1}{(\sigma-2)U^{\sigma-2}} \pm C
\frac{\left(1-(\sigma-1)U^{\sigma-1}\right)^
{\frac{\sigma-2}{\sigma-1}}}{U^{\sigma-2}},
\label{spec_sol}
\end{equation} 
where $C>0$ is an arbitrary constant. The + and - signs describe the
solution to the right and left of 
$\xi^*=-(\sigma-1)^{\frac{\sigma-2}{\sigma-1}}/(\sigma-2)$,
respectively. The special case $\sigma=4$ is shown in Fig.~\ref{alinhac_fig}.
The similarity solution (\ref{spec_sol}) is not smooth at its maximum;
rather, its first derivative behaves like 
$U_{\xi}\propto (\xi-\xi^*)^{1/(\sigma-2)}$. This can be understood from
the exact solution; in order for blow-up to occur at the same time
that a shock is formed, the initial profile must already have a maximum
with the same regularity as (\ref{spec_sol}). Thus, the situation
leading to (\ref{be_spec}) is a very special one, requiring very 
peculiar initial conditions. 

\subsubsection{Viscous pinch-off} 
\label{viscous}
As explained in section \ref{thin}, the pinch-off of a very viscous 
fluid is described by (\ref{hequ}), (\ref{vequ}), with $\rho = 0$,
but only for finite range of scales. 
The equations can be simplified considerably
by introducing \textit{Lagrangian variables}, i.e. writing all profiles
as a function of a particle label $s$. This means the particle is at 
position $z(s,t)$ at time $t$, and $z_t(s,t)$ is the velocity at 
time $t$. The jet profile can be obtained from 
$z_s=1/h^{2}(s,t)$, and (\ref{vequ}) becomes 
\begin{equation}
h_{t}(s,t)=\frac{1}{6}\left(1+\frac{C(t)}{h(s,t)}\right).
\label{visc}
\end{equation}
The typical velocity scale is $\gamma/\eta$, where $\gamma$ is 
the surface tension and $\eta$ is the viscosity; (\ref{visc})
has been made dimensionless accordingly. The time-dependent
constant of integration $C(t)$ has to be determined self-consistently.
Note that the self-similar form (\ref{ss}) is a solution of (\ref{visc}) for
$\alpha = 1$, and \textit{any} value of $\beta$; the exponent $\beta$
will be determined by the consistency condition (\ref{cons_cond}) below.

\begin{figure}[t]
\centering
\includegraphics[width=0.4\hsize]{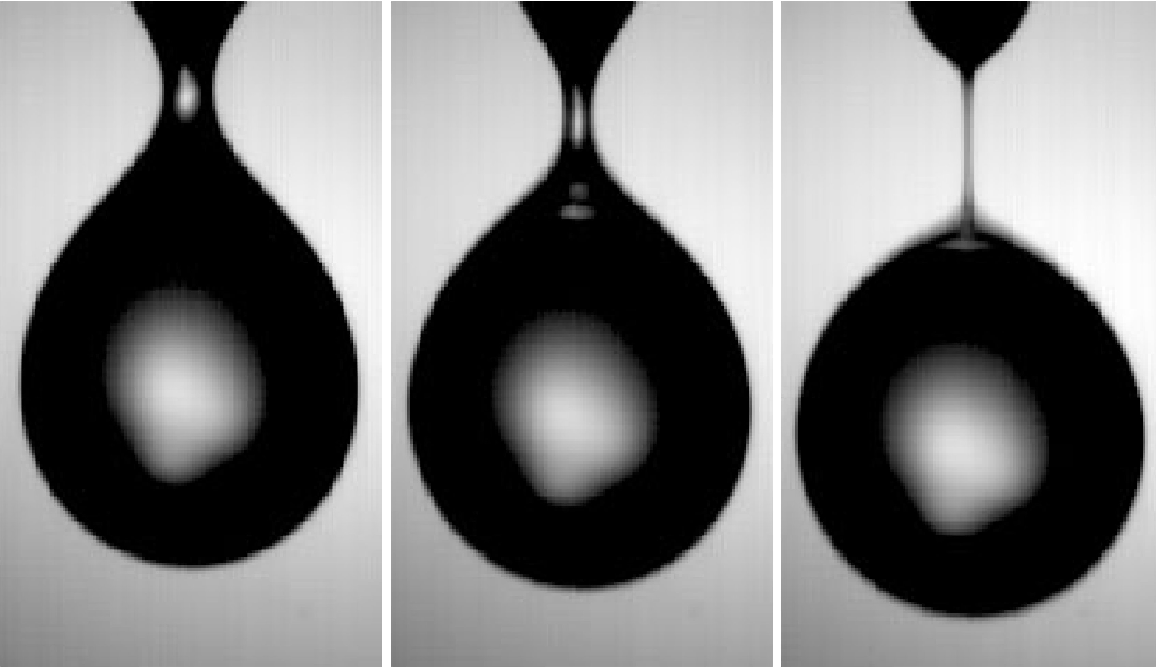}
\caption{A drop of viscous fluid falling from a pipette 1 mm in diameter
\protect\cite{RRR03}. Note the long neck. }
\label{rrr}
\end{figure}

Since $\alpha =1$, a scaling solution of (\ref{visc}) has the form
\begin{equation}
h^{-2}(s,t)=t^{\prime -2}f\left( \xi \right),\quad 
\mbox{with}\quad \xi =s'/t'^{\gamma}  \label{u}
\end{equation}
and 
\begin{equation}
C(t)=-C_0 t^{\prime }\ .  \label{paja21}
\end{equation}
The relationship with the exponent $\beta$ defined in (\ref{gen_sol})
is simply $\beta = \gamma-2$, as found from passing from Lagrangian
to Eulerian variables. Inserting (\ref{u}),(\ref{paja21}) 
into (\ref{visc}) we obtain
\begin{equation}
\frac{1}{\sqrt{f}}+3\left( \frac{2}{f}+\frac{\gamma \xi f_{\xi}}{f^{2}}%
\right) =C_0 ,  \label{d1}
\end{equation}
where $C_0$ is a constant. Imposing symmetry and regularity of $f$,
we expand $f(\xi)$ in the form
\begin{equation}
f_{i}(\xi)=R_0^{-2}+\xi ^{2i+2} + O(\xi^{2i+4})\ ,\ i=0,1,2,\dots
\label{cond}
\end{equation}
where we have normalised the coefficient of $\xi^{2i+2}$ to one. This 
is consistent, since any solution of (\ref{visc}) is only determined 
up to a scale factor. Instead, the axial scale is fixed by the 
initial conditions. 
The parameter $R_0$ is the rescaled minimum of the profile:
$h_m = R_0 t'$. 
Inserting (\ref{cond}) into (\ref{d1}), at order $\xi^{2i+2}$ one obtains
\begin{equation}
R_0=\frac{1}{12(\overline{\gamma} -1)}, 
\quad C_0 = \frac{1}{24}\frac{2\overline{\gamma }-1} {(
\overline{\gamma }-1)^{2}}  \label{min}
\end{equation}
where we have put $\overline{\gamma }=(i+1)\gamma$. 

Each choice of $i$ corresponds to one member in an infinite sequence 
of similarity solutions. Equation (\ref{d1}) can easily be
integrated in terms of $\ln\xi $ and $y=\sqrt{f}$:
\[
\int \frac{dy}{\left( \left( 1+6R_0\right) y^{3}-y^{2}-6R_0 y\right) }=
\frac{1}{6R_0\gamma }\ln \xi + \widetilde{C}=\frac{1}{6R_0\overline{\gamma
}} \ln \xi^{i+1}+\widetilde{C},
\]
with $\widetilde{C}$ an arbitrary constant. Computing the integral above we
obtain
\begin{equation}
y^{-\overline{\gamma}}\left( \left( 2\overline{\gamma}-1\right)y+1\right)
^{\overline{\gamma}-\frac{1}{2}}\left( 1-y\right)^{\frac{1}{2}}=\xi^{i+1},
\label{d3.1}
\end{equation}
which is an implicit equation for the i-th similarity profile
$y \equiv y_{i}(\xi) = \sqrt{f_{i}(\xi)}$. 

The value of the velocity $U_{\infty}$ at infinity must be a constant
to be consistent with boundary conditions. It can be found by 
integrating $z_{ts} = (h^{-2})_t = t'^{-3}(2f+\gamma\xi f_{\xi})$ from
zero to infinity:
\begin{equation}
U_{\infty}=\int_{0}^{\infty }z_{ts}ds = 
\frac{t'^{\gamma-3}}{3}
\int_0^{\infty}\left( \left( \frac{1}{24}\frac{2\overline{\gamma }-1} {(%
\overline{\gamma }-1)^{2}}\right) f^{2}-f^{\frac{3}{2}}\right) d\xi=0,
\label{cons_cond}
\end{equation}
where we have used (\ref{d1}). The above condition $U_{\infty}=0$, which 
ensures that $U_{\infty}$ does not diverge as $t'\rightarrow 0$, is
the equation which determines the exponent $\gamma$.
Taking the derivative of (\ref{d3.1}) we obtain
\[
(i+1)\xi ^{i}\frac{d\xi }{dy}=\frac{d}{dy} \left( y^{-\overline{\gamma}%
}\left(\left( 2\overline{\gamma }-1\right) y+ 1\right) ^{\overline{\gamma }-%
\frac{1}{2}}\left( 1-y\right) ^{\frac{1}{2}}\right) =
\]
\[
=-y^{-\overline{\gamma }-1} \left( 2y\overline{\gamma }-y+1\right) ^{\overline{%
\gamma }- \frac{3}{2}}\frac{\overline{\gamma }}{\sqrt{\left(1-y\right) }}
\]
which can be used to transform the integral in (\ref{cons_cond}) to
the variable $y$:
\begin{eqnarray}
&& K_{i}(\gamma )\equiv \frac{3U_{\infty}}{(12(\overline{\gamma}-1))^{3}}=
\frac{\overline{\gamma }}{i+1}
\int_0^1 \left(\left(\frac12\frac{2\overline{\gamma}-1
}{\overline{\gamma }-1}\right) y^4-y^3\right)\cdot  \nonumber \\
&& \left( y^{-\frac{i+1+\overline{\gamma }}{i+1}} 
\left( \left( 2\overline{\gamma }
-1\right)y+1\right)^{-\frac 12 \frac{2i-2\overline{\gamma}+3}{i+1}}\left(
1-y\right) ^{-\frac 12\frac{2i+1}{i+1}}\right) dy = 0.  \label{minne6}
\end{eqnarray}
The function $K_{i}(\gamma )$ may be written explicitly as
\begin{eqnarray}
&& K_{i}(\gamma ) =\gamma \frac{\Gamma \left( 4-\gamma \right) \Gamma\left(
\frac{1}{2i+2}\right) }{\Gamma \left( 4-\gamma+\frac{1}{2i+2}\right)}
\left(\frac{1}{2}\frac{(2i+2)\gamma -1}{(i+1)\gamma-1}\right)\cdot \nonumber \\
&& F\left( \frac{2i+3}{2i+2}-\gamma,4-\gamma ;
4-\gamma +\frac{1}{2i+2};1-(2i+2)\gamma\right) - 
\gamma \frac{\Gamma \left( 3-\gamma \right) \Gamma 
\left(\frac{1}{2i+2}\right)}{\Gamma\left(3-\gamma+\frac{1}{2i+2}\right) }\cdot
\nonumber \\
&& F\left( \frac{2i+3}{2i+2}-\gamma ,3-\gamma ;
3-\gamma + \frac{1}{2i+2};1-(2i+2)\gamma \right),  \label{kn}
\end{eqnarray}
where $F(a,b;c,z)$ is the hypergeometric function \cite{AS68}.
Roots of $\gamma_i$ are given in Table \ref{exponents}. 

To summarise, each exponent $\gamma_i$ corresponds to a new member 
$f_i(\xi)$ of an infinite hierarchy of similarity profiles, to be found from
(\ref{d3.1}). If one converts the Lagrangian
variables back to the original spatial variables, one obtains
\begin{equation}
h(x,t) = t'\phi^{(n)}_{St}\left(x'/t'^{\gamma-2}\right) .
\label{orig}
\end{equation}
Thus for $t'\rightarrow 0$ the typical radial scale $t'$
of the generic $i=0$ solution rapidly becomes smaller than the axial scale 
$t'^{0.175}$ (cf. Table \ref{exponents}). This explains the long necks
seen in Fig.~\ref{rrr}.

\begin{table}[tbp]
\begin{center}
\leavevmode
\begin{tabular}{ccc}
i & $\gamma_i$ & $R_0$   \\ \hline
0 & 2.1748 & 0.0709    \\
1 & 2.0454 & 0.0797   \\
2 & 2.0194 & 0.0817    \\
3 & 2.0105 & 0.0825    \\
4 & 2.0065 & 0.0828   \\
5 & 2.0044 & 0.0832  \\
&  &  
\end{tabular}
\end{center}
\caption{A list of exponents, found from $K_i(\protect\gamma) = 0$
using MAPLE, with $K_i$ given by (\protect\ref{kn}). The number $2i+2$ 
gives the smallest
non-vanishing power in a series expansion of the corresponding similarity
solution around the origin. Only the solution with $i=0$ is stable. The
rescaled minimum radius is found from (\protect\ref{min}).}
\label{exponents}
\end{table}

\subsubsection{More examples} 
\label{more}
Other recent examples for scaling of the second kind have been
observed for the breakup of a two-dimensional sheet with
surface tension. In a shallow-water approximation, which is 
justified for a description of breakup, the equations read
\cite{BT07}
\begin{equation} \label{bt_sys}
h_t + (hu)_x =0, \quad u_t + uu_x = h_{xxx}
\end{equation}
after appropriate rescaling. Local similarity solutions can
be found in the form 
\begin{equation} \label{bt_sol}
h(x,t) = t'^{4\beta-2}H(\eta), \quad 
u(x,t) = t'^{\beta-1}U(\eta), 
\end{equation}
where $\eta = x'/t'^{\beta}$. The exponent $\beta$ is not
determined by dimensional analysis. Instead, it must be found
from a solvability condition on the nonlinear system
of equations for the similarity functions $H,U$. 

The result of the numerical calculation is \cite{BT07}
$\beta = 0.6869\pm 0.0003$, which is curiously close 
to $\beta = 2/3$, which is the value that had been conjectured 
earlier \cite{DGKZ93}, but contains a small correction. 
The value $\beta = 2/3$ comes
out if both length scales in the longitudinal and transversal 
directions are assumed to be the same, implying that 
$4\beta - 2 = \beta$. This is a natural expectation 
for problems governed by Laplace's equation, such as 
inviscid, irrotational flow \cite{DHL98}, and indeed is 
observed for three-dimensional drop breakup \cite{CS97,DHL98}.
However, in present case, even if the full two-dimensional
irrotational flow equations are used, $\beta \ne 2/3$. 

Other physical problems which frequently involve anomalous scaling
exponents are strong explosions on one hand, and collapse of 
particles or gases into a singular state on the other. These 
types of problems have been reviewed in great detail in a number
of textbooks and articles \cite{BZ72,Sachdev,Sedov,B96}, but
continue to attract a great deal of attention. As with many
other singular problems, the type of scaling depends on the 
details of the underlying physics, and scaling of both the first
and second kind is observed. For example, the radius of a shock
wave resulting from a strong explosion can be calculated from 
dimensional analysis to be $r_s\propto t^{2/5}$ \cite{Barenblatt}.
However, in the seemingly analogous case of a strong {\it implosion}, an 
anomalous exponent is observed, which moreover depends on the 
parameters of the problem \cite{G42,LL84a}. Cases were collapse 
and shock formation coincide were given by \cite{BW98} 
(similar to section \ref{shock} above). 
In a somewhat different context, anomalous scaling
is observed in model calculations for the collapse of 
self-gravitating particles \cite{CS04} and Bose-Einstein condensates
\cite{LLPR01}. It is important to remember that these examples come
from {\it kinetic} equations describing the stochastic collision of
waves or particles, and hence involving nonlocal collision operators.
However, the kinetic equations appear to be closely related to 
certain PDE problems \cite{JPR06}, which are analogous to other 
evolution equations studied in this article. 

\subsection{Stability of fixed points}
\label{stable}
Self-similar solutions correspond to fixed points of the dynamical 
system (\ref{ds}), whose stability we now investigate by linearising 
around the fixed point. We explain the 
situation for the example of section \ref{sub:first} in more detail,
for which the transformation reads
\begin{equation}
h(x,t)=t'^{1/4} H(\xi,\tau),
\label{ssmean}
\end{equation}
where $\tau = -\ln(t')$. The similarity form of (\ref{sd}) becomes 
\begin{equation}
H_{\tau}=\frac{1}{4}(H-\xi H_{\xi})+\frac{1}{H}
\left[\frac{H}{(1+H_{\xi}^2)^{1/2}}\kappa_{\xi}\right]_{\xi},
\label{mean_dyn}
\end{equation}
which reduces to (\ref{sdsim}) if the left hand side is set to 
zero. To assure matching of (\ref{mean_dyn})
to the outer solution, we have to require that (\ref{ssmean}) is 
to leading order time-independent as $\xi$ is large, which leads
to the boundary condition 
\begin{equation}
H_{\tau} - (H-\xi H_{\xi})/4\rightarrow 0 \quad \mbox{for}
\quad |\xi| \rightarrow \infty .
\label{bcsd}
\end{equation}
This is the natural extension of (\ref{bound}) to the time-dependent case.

Next we linearise around any one of the similarity solutions
$\overline{H}(\xi)=H_i(\xi)$ listed in Table \ref{series}, as
described in the Introduction. The stability is controlled by
eigenvalues of the eigenvalue equation (\ref{mean_lyn}).
Inserting the eigensolution (\ref{eigen_exp}) into (\ref{bcsd}) 
one finds that $P_j$ must grow at infinity like 
\begin{equation}
P_j(\xi) \propto \xi^{1-4\nu_j}. 
\label{eigen_grow}
\end{equation}
Similarly, the growth condition for the general case of a similarity 
solution of the form (\ref{ss}) is 
\begin{equation}
P_j(\xi) \propto \xi^{\frac{\alpha-\nu_j}{\beta}}. 
\label{eigen_grow_gen}
\end{equation}

If the similarity solution $\overline{H}(\xi)$ is to be stable,
the real part of the eigenvalues of ${\cal L}$ must be negative. 
However, there are always two positive eigenvalues, which are
related to the invariance of the equation of motion (\ref{sd}) 
under translations in space and time, as noted by \cite{FK92,VGH91}.
Namely, for any $\epsilon$, the translated similarity solution
\begin{equation}
h^{(\epsilon)}(x,t) = t'^{1/4}\overline{H}(\frac{x'+\epsilon}{t'^{1/4}}) 
\label{trans}
\end{equation}
is an equally good self-similar solution of (\ref{sd}), and
thus of (\ref{mean_dyn}). 
In particular, we can expand (\ref{trans}) to lowest order in 
$\epsilon$, and find that 
\begin{equation}
H^{(\epsilon)}(\xi,\tau) = \overline{H}(\xi) + 
\epsilon e^{\beta\tau} \overline{H}_{\xi}(\xi) + O(\epsilon^2),
\label{trans_sim}
\end{equation}
where the linear term is a solution of (\ref{sd_lin}). 

Thus
\begin{equation}
\left(e^{\beta\tau} \overline{H}_{\xi}\right)_{\tau}  = 
e^{\beta\tau} \beta \overline{H}_{\xi} = 
e^{\beta\tau} {\cal L} \overline{H}_{\xi} .
\label{f_exp}
\end{equation}
But this means that $\nu_x = \beta\equiv 1/4$ is an eigenvalue of
${\cal L}$ with eigenfunction $\overline{H}_{\xi}(\xi)$. Similarly, considering
the transformation $t\rightarrow t + \epsilon$, one finds a second 
positive eigenvalue $\nu_t = 1$, with eigenfunction
$\xi\overline{H}_{\xi}$. However, these two positive eigenvalues {\it do not}
correspond to instability. Instead, the meaning of these 
eigenvalues is that upon perturbing the similarity solution, the singularity
time as well as the position of the singularity will change. Thus if
the coordinate system is not adjusted accordingly, it looks as if the 
solution would flow away from the fixed point. If, on the other hand,
the solution is represented relative to the perturbed values of 
$x_0$ and $t_0$, the eigenvalues $\nu_x$ and $\nu_t$ will not
appear. 

The eigenvalue problem (\ref{mean_lyn}) was studied numerically 
in \cite{BBW98}. It was found that each similarity solution 
$\overline{H}_i$ has exactly $2i$ positive real eigenvalues,
{\it disregarding} $\nu_x,\nu_t$. The result is that the linearisation 
around the ``ground state'' solution $\overline{H}_0$ only has 
negative eigenvalues while {\it all} the other solutions have at 
least one other
positive eigenvalue. This means that $\overline{H}_0$ is the only 
similarity solution that can be observed, all other solutions 
are unstable. Close to the fixed point, the approach to $\overline{H}_0$ 
will be dominated by the largest negative eigenvalue $\nu_1$:
\begin{equation}
h(x,t)= t'^{1/4} \left[\overline{H}(\xi) + 
\epsilon t'^{-\nu_1} P_1(\xi)\right]. 
\label{approach}
\end{equation}
For large arguments, the point $\xi_{cr}$ where the correction becomes
comparable to the similarity solution is 
$\xi \sim \epsilon t'^{-\nu_1} \xi^{1-4\nu_1}$, and thus 
$\xi_{cr} \sim t'^{-1/4}$. This means that the region of validity 
of $\overline{H}(\xi)$ {\it expands} in similarity variables, and 
is constant in real space. 
This rapid convergence is reflected by the numerical results reported in 
Fig. \ref{sdfig}. More formally, one can say that for
any $\epsilon$ there is a $\delta$ such that 
\begin{equation}
\left|h(x,t) - t'^{1/4}\overline{H}(\xi)\right| \le \epsilon 
\label{conv1}
\end{equation}
if $|x'|\le \delta$ {\it uniformly} as $t' \rightarrow 0$. 

We suspect that the situation described above is more general:
the ground state is stable, while each following profile has 
a number of additional eigenvalues. In the case of the sequence
of profiles $\overline{H}_i$ of (\ref{sdsim}), two new positive 
eigenvalues appear for each new profile, corresponding to a 
symmetric and an antisymmetric eigenfunction. 
Below we give two more examples of the same scenario,
for which we are able to give a simple geometrical interpretation 
for the appearance of two additional positive eigenvalues at each 
stage of the hierarchy of similarity solutions. The simplest case is
that of shock wave formation (cf. section \ref{shock}), for which
everything can be worked out analytically.

The dynamical system corresponding to the self-similar solution
(\ref{be_scal}) is 
\begin{equation}
U_{\tau}- \alpha U + \left(1 + \alpha\right)\xi U_{\xi} + UU_{\xi} = 0, 
\label{be_ds}
\end{equation} 
and so the eigenvalue equation for perturbations $P$ around the 
base profile $\overline{U}_i$ becomes 
\begin{equation}
(\alpha_i - \nu)P - (1 + \alpha_i)\xi P_{\xi} - P(\overline{U}_i)_{\xi} -
P_{\xi}\overline{U}_i = 0, \quad i=0,1,\dots
\label{be_ev}
\end{equation} 
Here $\overline{U}_i$ is the ith similarity function defined by 
(\ref{be_impl}) for the exponents $\alpha_i$ as given by 
(\ref{be_alpha}).

The eigenvalue equation (\ref{be_ev}) is solved easily by 
transforming from the variable $\xi$ to the variable $\overline{U}$,
using (\ref{be_impl}):
\begin{equation}
P\left[(\alpha_i - \nu)(1+(2i+3)\overline{U}_i^{2i+2})+1\right] =
\frac{\partial P}{\partial\overline{U}}
\left[\alpha_i\overline{U}_i + (1+\alpha_i)\overline{U}_i^{2i+3}\right],
\label{be_ev_tr}
\end{equation} 
with solution 
\begin{equation}
P = \frac{\overline{U}_i^{3+2i-2\nu (i+1)}}{1+(2i+3)\overline{U}_i^{2i+2}}.
\label{be_ev_sol}
\end{equation} 
The exponent $3+2i-2\nu (i+1)$ must be an integer for (\ref{be_ev_sol})
to be regular at the origin, so the eigenvalues are 
\begin{equation}
\nu_j = \frac{2i+4-j}{2i+2}, \quad j = 1,2,\dots
\label{be_evs}
\end{equation} 
As usual, the eigensolutions are alternating between even and odd. 
However, we are interested in the {\it first} instance, given 
by (\ref{first}), at which a shock forms. This implies that the
second derivative of the profile must vanish at the location of 
the shock, and the amplitude of the $j=3$ perturbation must be 
exactly zero. 

Thus for $i=0$ the remaining eigenvalues are $\nu = 3/2,1,0,-1/2,\dots$;
the first two are the eigenvalues $\nu_x=\beta=1+\alpha$ and 
$\nu_t=1$ found above. The vanishing eigenvalue occurs because 
there is a family of solutions parameterised by the coefficient 
$C$ in (\ref{be_impl}). All the other eigenvalues are negative,
which shows that the similarity solution (\ref{be_alpha1}) is 
stable. In the same vein, for $\alpha_1=1/4$ there are two more
positive exponents: $\nu=5/4,1,1/2,1/4$, so the solution must
be unstable. The same is of course true for all higher order
solutions. Thus in conclusion the ground state solution 
$\overline{U}_0$ given by (\ref{be_alpha1}) is the only observable 
form of shock formation. The same conclusion was reached in \cite{PLGG08}
by a stability analysis based on catastrophe theory. 

The sequence of profiles for viscous pinch-off, found in section 
\ref{second}, suggests a simple mechanism for the fact that two new
unstable directions appear with each new similarity profile
of higher order. In fact, the argument is strikingly similar to that 
given for shock formation. 
Differentiating (\ref{visc}) with respect to 
$s$ one finds that a local minimum point $s_{min}$ remains a 
minimum. Thus the local time evolution of the profile can be written as
\begin{equation}
h(s,t) = h_m + \sum_{j=2}^{\infty} B_j(t)s'^j. 
\label{init1}
\end{equation}
For generic initial data $B_2(0) \ne 0$, so there is no reason why
$B_2$ should vanish at the singular time, which means that the self-similar 
solution $f_0$ will develop, which has a quadratic minimum. This 
situation is structurally stable, so one
expects the eigenvalues of the linearisation to be negative. 
If however the coefficients $B_j(0)$ are zero for $j=2,\dots 2n-1$, 
they will remain zero for all times. Namely, if the first $k$ $s$-derivatives 
of $h$ vanish, one has 
\begin{equation}
\partial_s^j h_{t} = -\frac{C\partial_s^j h}{h^2}, \quad 
j = 1,\dots,k, 
\label{diff_visc}
\end{equation}
so the first $k$ derivatives will remain zero. Thus to find the 
similarity profile with $i=1$, one needs $B_2(0) = B_3(0) = 0$
as an initial condition. This is a non-generic situation,
and a slight perturbation will make $B_2$ and $B_3$ nonzero. 
In other words, there are two unstable directions, which take 
the solution away from $f_1(\xi)$, as defined by (\ref{cond}). 
In the general case, the linearisation around $f_i(\xi)$ will
have $2i$ positive eigenvalues (apart from the trivial ones).
Extensive numerical simulations of drop pinch-off in the 
inertial-surface tension-viscous regime (cf. section \ref{thin})
suggests that the the hierarchy of similarity solutions again has 
similar properties in this case as well, although stability has not
been studied theoretically. The ground-state profile is stable, 
while all the others are unstable \cite{Edrop05}. 
Even when using a higher-order similarity solution as an initial
condition, it is immediately destabilised, and converges onto the
ground state solution \cite{BLS96}. 

\section{Centre manifold}
\label{sec:centre}

In section \ref{fixed} we described the generic situation that the behaviour
of a similarity solution is determined by the linearisation around it. In
the case of a stable fixed point, convergence is exponentially fast, 
and the observed behaviour is essentially that of the fixed point. 
In this section, we describe a variety different cases where the 
the dynamics is slow. In all cases we are able to associate this 
slow dynamics with a fixed point in the appropriate variable(s),
around which the eigenvalues vanish. Instead, higher-order non-linear 
terms have to be taken into account, and the slow approach to 
the fixed point is determined by a low-dimensional dynamical system. 

We consider essentially two different cases: 
\begin{itemize}
\item[(a)] The dynamical system (\ref{ds}) possesses a fixed point
$H_0(\xi)$, which has a {\it vanishing} eigenvalue, with corresponding 
eigenfunction $\psi(\xi)$. The dynamics in the slow direction $\psi$
is described by a nonlinear equation for the amplitude $a(\tau)$, 
which varies on a logarithmic time scale:
\begin{equation}
h = t'^{\alpha}\left[H_0(\xi) + a(\tau)\psi(\xi)\right], 
\quad \xi=x'/t'^{\beta} .
\label{model1}
\end{equation}
\item[(b)] The dynamical system does not possess a fixed point,
but has a solution of a slightly more general form:
\begin{equation}
h = h_0(\tau)H(\xi), \quad \xi=x'/W(\tau),
\label{model2}
\end{equation}
where $h_0$ and $W$ are not necessarily power laws. To expand
about a fixed point, we define the generalised exponents 
\begin{equation}
\alpha = -\partial_{\tau}h_0/h_0, \quad 
\beta = -\partial_{\tau}W/W
\label{model3}
\end{equation}
which now depend on time. In the case of a type-I similarity 
solution, this reduces to the usual definition of the exponent. 
In the cases considered below, one derives a finite dimensional 
dynamical system for the exponents $\alpha,\beta$ (potentially 
including other, similarly defined scale factors). Once more,
the exponents vary on a logarithmic time scale, which can be 
understood from the fact that the dynamical system possesses 
a fixed point with vanishing eigenvalues. 
\end{itemize}

Zero eigenvalues can also be associated to symmetries of the
singularity, like rotational or translational symmetries, which lead
to the existence of a continuum of similarity solutions. Another 
example, which concerns the dynamics inside the singular object itself,
is wave steepening as described by (\ref{be_impl}) above. As seen
from (\ref{be_evs}), there indeed is a vanishing eigenvalue associated
with this continuum of solutions. Below we will not be concerned with
this case, but only consider approach to the singularity starting 
from nonsingular solutions. 

\subsection{Quadratic non-linearity: geometric evolution and
reaction-diffusion equations}

The appearance of this type of nonlinearity is characteristic for various
nonlinear parabolic equations and systems. The blow-up behaviour is
characterised by the presence of logarithmic corrections in the similarity
profiles.

\subsubsection{Geometric evolution equations: Mean curvature and Ricci flows}
\label{quad}
\begin{figure}
\centering
\centerline{\includegraphics[width=0.4\linewidth]{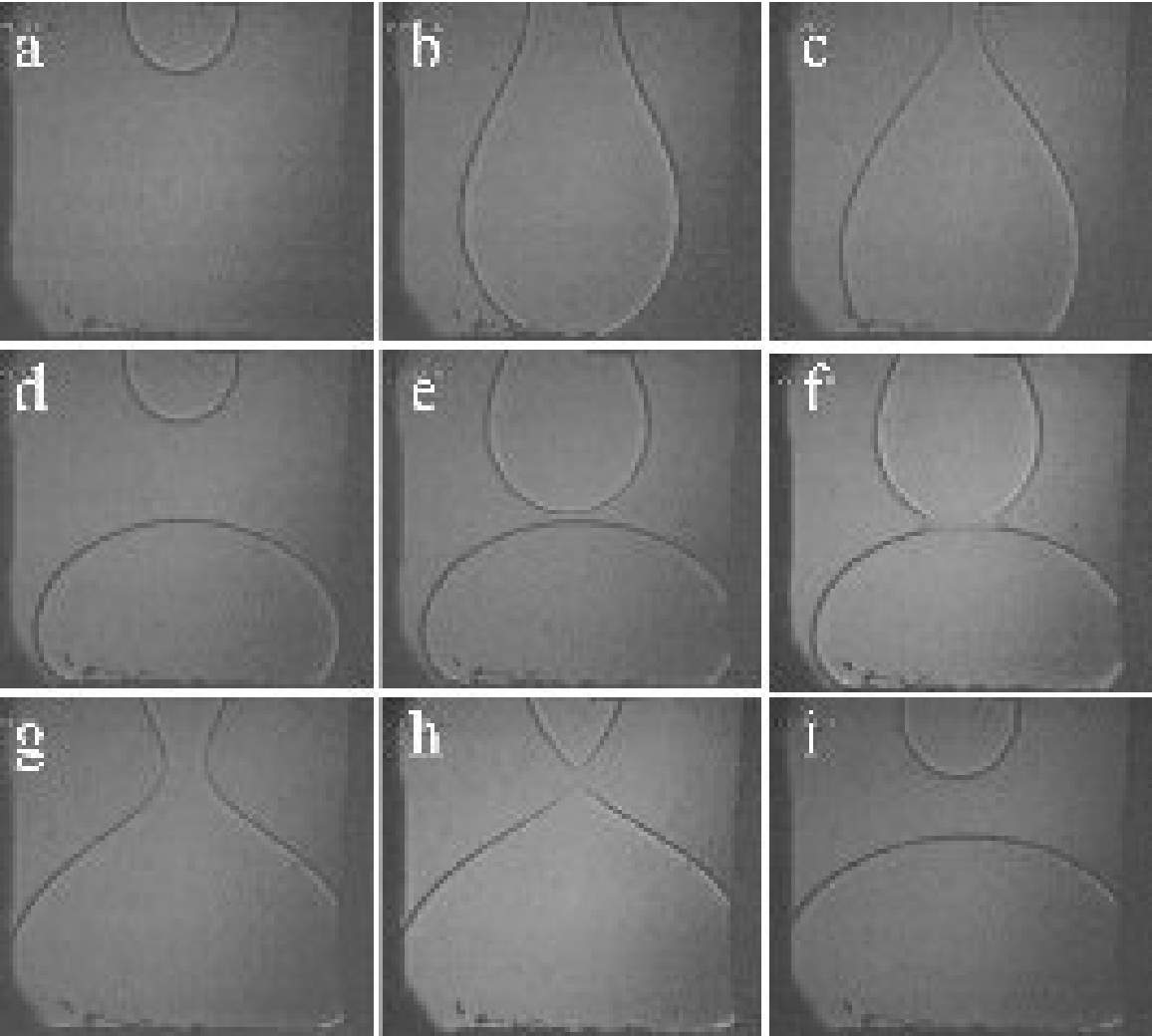}}
\vspace{0.5 cm}
\caption{Nine images (of width 3.5 mm) showing how 
a $^3$He crystal ``flows'' down from the upper part of a cryogenic 
cell into its lower part \cite{Ishiguro}. The recording takes a 
few minutes, the temperature is 0.32 K. 
11 mK. The crystal first ``drips'' down, so that a crystalline 
``drop'' forms at the bottom (a to c); then a second drop appears 
(d) and comes into contact with the first one (e); coalescence 
is observed (f) and subsequently breakup occurs (h). }
\label{f2}
\end{figure}

Axisymmetric motion by mean curvature in three spatial
dimensions is described by the equation 
\begin{equation}
h_t= \left(\frac{h_{xx}}{1+h_x^2} - \frac{1}{h}\right),
\label{mc}
\end{equation}
where $h(x,t)$ is the radius of the moving free surface. 
A very good physical realization of (\ref{mc}) is the melting
and freezing of a $^3$He crystal, driven by surface tension \cite{IGRBE07},
see Fig.~\ref{f2}. As before, the time scale $t$ has been 
chosen such that the diffusion constant, which sets the rate
of motion, is normalised to one. 
A possible boundary condition for the problem is 
that $h(0,t) = h(L,t) = R$, where $R$ is some prescribed 
radius. For certain initial conditions $h(x,0)\equiv h_0(x)$ 
the interface will become singular at some time $t_0$, 
at which $h(x_0,t_0)=0$ and the curvature blows up. The 
moment of blow-up is shown in panel h of Fig.~\ref{f2}, 
for example.  

Inserting the self-similar solution (\ref{ss}) into (\ref{mc}),
one finds a balance for $\alpha = \beta = 1/2$. The corresponding 
similarity equation is
\begin{equation}
-\frac{\phi}{2}+\xi\frac{\phi_{\xi}}{2} = 
\left(\frac{\phi_{\xi\xi}}{1+\phi_{\xi}^2} - \frac{1}{\phi}\right), \quad
\xi = \frac{x'}{t'^{1/2}}. 
\label{mean_sim}
\end{equation}
One solution of (\ref{mean_sim}) is the constant 
solution $\phi(\xi) = \sqrt{2}$. 
Another potential solution is one that grows linearly at infinity,
to ensure matching onto a time-independent outer solution. 
However, it can be shown that no solution to (\ref{mean_sim}),
which also grows linearly at infinity, exists \cite{AAG95,Huis93}.
Our analysis below follows the rigorous work in \cite{AV97}, 
demonstrating type-II self-similarity. In addition, we now show
how the description of the dynamical system can be carried out
to arbitrary order. 

The relevant solution is thus the constant solution, but which of 
course does not match onto a time-independent outer solution. We thus 
write the solution as 
\begin{equation}
h(x,t)=t'^{1/2}\left[\sqrt{2}+g(\xi,\tau)\right],
\label{blmc}
\end{equation}
with $\tau = -\ln(t')$ as usual. 
The equation for $g$ is then 
\begin{equation}
g_{\tau}=g-\frac{\xi g_{\xi}}{2}+\frac{g_{\xi\xi}}{1+g_{\xi}^2} - 
\frac{g^2}{2^{3/2}+2g},
\label{gequ}
\end{equation}
which we solve by expanding into eigenfunctions of the linear
part of the operator 
\begin{equation}
{\cal L}g = g-\xi g_{\xi}/2+g_{\xi\xi}.
\label{ling}
\end{equation}

It is easily confirmed that 
\begin{equation}
{\cal L}H_{2i}(\xi/2) = \nu_i H_{2i}(\xi/2), \quad i=0,1,\dots,
\label{eigen}
\end{equation}
where $H_n$ is the n-th Hermite polynomial \cite{AS68}:
\begin{equation}
H_n(y) = (-1)^ne^{y^2}\frac{d^n}{dy^n}e^{-y^2},
\label{Hermite}
\end{equation}
and $\nu_i = 1-i$. 
Thus the first eigenvalue is $\nu_0=1$, which corresponds to 
the positive eigenvalue $\nu_t$ coming from the arbitrary choice of $t_0$. 
The other positive eigenvalue eigenvalue $\nu_x$ does not appear, since
we have chosen to look at symmetric solutions, breaking translational
invariance. However, the largest non-trivial eigenvalue $\nu_1$ is 
zero, and the linear part of (\ref{gequ}) becomes
\begin{equation}
\frac{\partial a_i}{\partial\tau}=(1-i)a_i, \quad i=0,1,\dots.
\label{linai}
\end{equation}
Thus all perturbations with $i>1$ decay, but to investigate the approach 
of the cylindrical solution, one must include nonlinear terms in the 
equation for $a_1$. 

If we write
\begin{equation}
g(\xi,\tau) = \sum_{i=1}^{\infty} a_i(\tau)H_{2i}(\xi/2),
\label{gexp}
\end{equation}
the equation for $a_1$ becomes
\begin{equation}
\frac{d a_1}{d\tau} = -2^{3/2}a_1^2 + O(a_1a_j),
\label{a1}
\end{equation}
whose solution is 
\begin{equation}
a_1 = 1/(2^{3/2}\tau).
\label{a1_r}
\end{equation}

Thus instead of the expected
exponential convergence onto the fixed point, the approach is only 
algebraic. Since all other eigenvalues are negative, the $\tau$-dependence
of the $a_i$ is slaved by the dynamics of $a_1$. Namely, as we will see 
below, $a_j=O(\tau^{-j})$, so corrections to (\ref{a1}) are of
higher order. To summarise, the leading-order
behaviour of (\ref{mc}) is given by 
\begin{equation}
h(x,t)=t'^{1/2}\left[\sqrt{2}+a_1(\tau)H_2(\xi)\right],
\label{h_lead}
\end{equation}
as was proven by \cite{AV97}. 

Now  we compute the specific form of the higher-order corrections
to (\ref{h_lead}), which have not been worked out explicitly before. 
If one linearises around (\ref{a1_r}), putting $a_1 = a_1^{(0)} + \epsilon_1$,
one finds 
\begin{equation}
\frac{d \epsilon_1}{d\tau} = -\frac{2}{\tau}\epsilon_1 + \mbox{other terms}.
\label{a1_pert}
\end{equation}
This means that the coefficient $A$ of $\epsilon_1 = A/\tau^2$ remains
undetermined, and a simple expansion of $a_i$ in powers of $\tau^{-1}$
yields an indeterminate system. Instead, at quadratic order, a term of
the form $\epsilon_1 = A\ln\tau/\tau^2$ is needed. Fortunately, this is 
the only place in the system of nonlinear equations for $a_i$ where
such an indeterminacy occurs. Thus all logarithmic dependencies can be 
traced, leading to the general ansatz
\begin{equation}
a_i^{(n)} = \frac{\delta_i}{\tau^i} + \sum_{k=i+1}^n
\sum_{l=0}^{k-i}\frac{(\ln\tau)^l}{\tau^k}\delta_{lki},
\label{ai_gen}
\end{equation}
where $\delta_i$ and $\delta_{lki}$ are coefficients to be determined. 
The index $n$ is the order of the truncation. 

The coefficients can now be found recursively by considering 
terms of successively higher order in $\tau^{-1}$ in the first
equation:

\numparts
\begin{eqnarray}
&& \frac{d a_1}{d\tau} = -2^{3/2}a_1^2 - 24\sqrt{2}a_1a_2+22a_1^3 -\nonumber\\
    \label{ai_sys1}
&& 272\sqrt{2}a_1^4-191\sqrt{2}a_2^2+192a_1^2a_2 \\
\label{ai_sys2}
&& \frac{d a_2}{d\tau} = -a_2-\sqrt{2}/4a_1^2+6a_1^3-8\sqrt{2}a_1a_2 .
\end{eqnarray}
\endnumparts
The next two orders will involve the next coefficient $a_3$. 
From (\ref{ai_sys1}) and (\ref{ai_sys2}), one first finds 
$\delta_{121}$ and $\delta_2$, by considering $O(\tau^{-3})$ and 
$O(\tau^{-2})$, respectively. Then, at order $O(\tau^{-(n+1)})$ 
in the first equation, where $n = 3$, one finds all remaining 
coefficients $\delta_{lki}$ in the expansion (\ref{ai_gen}) up
to $k=n$. At each order in $\tau^{-1}$, there is of course a 
series expansion in $\ln\tau$ which determines all the coefficients. 

We constructed a MAPLE program to compute all the coefficients up
to arbitrarily high order (10th, say). Up to third order in $\tau^{-1}$
the result is:
\numparts
\begin{eqnarray}
&& a_1 = 1/4\,{\frac {\sqrt {2}}{\tau}}+{\frac {17}{16}}\,{\frac {\ln  \left(
\tau \right) \sqrt {2}}{{\tau}^{2}}}-{\frac {73}{16}}\,{\frac {\sqrt {
2}}{{\tau}^{3}}}+ \nonumber \\
&& {\frac {867}{128}}\,{\frac {\ln  \left( \tau \right)
\sqrt {2}}{{\tau}^{3}}}-{\frac {289}{128}}\,{\frac { \left( \ln
 \left( \tau \right)  \right) ^{2}\sqrt {2}}{{\tau}^{3}}} \\ 
&& a_2=-1/32\,{\frac {\sqrt {2}}{{\tau}^{2}}}+{\frac {5}{16}}\,{\frac {\sqrt
{2}}{{\tau}^{3}}}-{\frac {17}{64}}\,{\frac {\ln  \left( \tau \right) 
\sqrt {2}}{{\tau}^{3}}},
\label{ai_res}
\end{eqnarray}
\endnumparts
and thus $h(x,t)$ becomes
\begin{equation}
h(x,t)=t'^{1/2}\left[\sqrt{2}+a_1(\tau)\left(-2+\xi^2\right)
+ a_2(\tau)\left(12-12\xi^2+\xi^4\right)\right],
\label{h_res}
\end{equation}
from which one of course immediately finds the minimum. 
To second order, the result is 
\begin{equation}
h_m=(2 t')^{1/2}\left[1 - \frac{1}{2\tau}
-\frac{3+17\ln\tau}{8\tau^2} \right].
\label{hmin_res}
\end{equation}

\begin{figure}
\centering
\includegraphics[width=0.5\hsize]{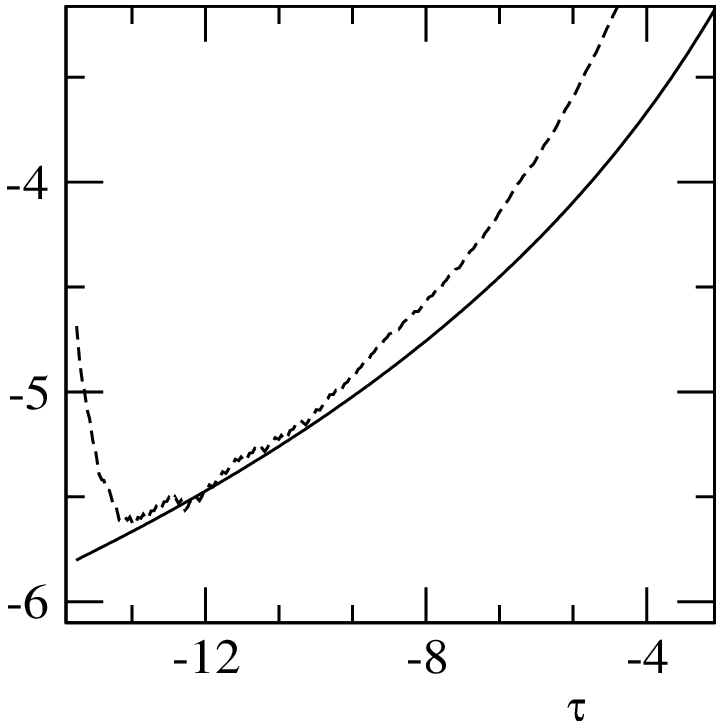}
\caption{\label{hminfig} 
A plot of $\left[h_m/\sqrt{2t'}-1+1/(2\tau)\right]\tau^2$ 
(dashed line) and $\tau_0/2 - (3+17\ln(\tau+\tau_0)/8)$ (full
line) with $\tau_0 = 4.56$. 
   }
\end{figure}
First, the presence of logarithms implies
that there is some dependence on initial conditions built into
the description. The reason is that the argument inside the logarithm
needs to be non-dimensionalised using some ``external'' time scale. 
More formally, any change in time scale $\tilde{t} = t/t_0$ leads to
an identical equation if also lengths are rescaled according to 
$\tilde{h} = h/\sqrt{t_0}$. This leaves the prefactor in 
(\ref{hmin_res}) invariant, but adds an arbitrary constant $\tau_0$ 
to $\tau$. This is illustrated by comparing to a numerical simulation
of the mean curvature equation (\ref{mc}) close to the point
of breakup, see Fig.~\ref{hminfig}. Namely, we subtract the analytical result
(\ref{hmin_res}) from the numerical solution $h_m/(2\sqrt{t'})$ 
and multiply by $\tau^2$. As seen in Fig.\ref{hminfig}, the remainder 
is varying slowly over 12 decades in $t'$. If the constant $\tau_0$ is 
adjusted, this small variation is seen to be consistent with the 
logarithmic dependence predicted by (\ref{hmin_res}).

The second important point is that convergence in space is no longer
uniform as implied by (\ref{conv1}) for the case of type I self-similarity.
Namely, to leading order the pinching solution 
is a cylinder. For this to be a good approximation, one has to require 
that the correction is small: $\xi^2/\tau \ll 1$. Thus corrections 
become important beyond $\xi_{cr} \sim \tau$, which, in view of the 
logarithmic growth of $\tau$, implies convergence in a constant region
{\it in similarity variables only}. As shown in \cite{IGRBE07}, the
slow convergence toward the self-similar behaviour has important 
consequences for a comparison to experimental data. 

Mean curvature flow is also an example of a broader class of problems called
generically "geometric evolution equations". These are evolution equations
intended to gain topological insight by flowing geometrical objects 
(such as metric or curvature) towards easily recognisable objects such 
as constant or positive curvature manifolds. The most remarkable 
example is the so called Ricci flow, introduced in \cite{H82},
which is the essential
tool in the recent proof of the geometrisation conjecture (including
Poincar\'e's conjecture as a consequence) by Grigori Perelman. 

Namely, Poincar\'{e}'s conjecture states that every simply connected closed
3-manifold is homeomorphic to the 3-sphere. Being homeomorphic means that
both are topologically equivalent and can be transformed one into the other
through continuous mappings. Such mappings can be obtained from the flow
associated to an evolutionary PDE involving fundamental geometrical
properties of the manifold.
Thurston's geometrisation conjecture is a generalisation of Poincar\'{e}'s
conjecture to general 3-manifolds and states that compact 3-manifolds can be
decomposed into submanifolds that have basic geometric structures.

Perelman sketched a proof of the full geometrisation conjecture in
2003 using Ricci flow with surgery \cite{P03}. Starting with an initial 
3-manifold, one deforms it in time according to the solutions of the Ricci 
flow PDE (\ref{Ricci_g}) we consider below. Since the flow is continuous, 
the different manifolds obtained during the evolution will be homeomorphic 
to the initial one. The problem is in the fact that Ricci flow develops 
singularities in finite time, one of which we describe below. 
One would like to get over this difficulty by devising a mechanism of 
continuation of solutions beyond the singularity, making sure that 
such a mechanism controls the topological changes leading to a decomposition 
into submanifolds, whose structure is given by Thurston's 
geometrisation conjecture. Perelman obtained essential
information on how singularities are like, essentially three dimensional
cylinders made out of spheres stretched out along a line, so that he could
develop the correct continuation (also called ``surgery'') procedure and
continue the flow up to a final stage consisting of the elementary
geometrical objects in Thurston's conjecture.

Ricci flow is defined by the equation
\begin{equation}
\frac{\partial g_{ij}}{\partial t}=-2R_{ij}
\label{Ricci_g}
\end{equation}
for a Riemannian metric $g_{ij}$, where $R_{ij}$ is the Ricci curvature
tensor. The Ricci tensor involves second derivatives of the curvature and
terms that are quadratic in the curvature. Hence, there is the potential for
singularity formation and singularities are, in fact, formed. As Perelman
poses it, the most natural way to form a singularity in finite time is by
pinching an almost round cylindrical neck. The structure of this kind of
singularity has been studied in \cite{AK08}. By
writing the metric of a $(n+1)$-dimensional cylinder as
\begin{equation}
g=ds^{2}+\psi ^{2}g_{can}\ ,
\label{Ricci_metric}
\end{equation}
where $g_{can}$ is the canonical metric of radius one in the $n-$sphere 
$S^{n}$, $\psi (s,t)$ is the radius of the hypersurface $\left\{ s\right\}
\times S^{n}$ at time $t$ and $s$ is the arclength parameter of the
generatrix of the cylinder. 

The equation for $\psi$ then becomes 
\begin{equation}
\psi_t = \psi_{ss} - 
\frac{(n-1)(1-\psi_s^2)}{\psi}. 
\label{Ricci}
\end{equation}
In \cite{AK08} it is shown that for $n>1$ the solution close to 
the singularity admits a representation that resembles the one
obtained for mean curvature flow:
\begin{equation}
\psi(s,t)=\frac{1}{2^{\frac{1}{2}}(n-1)^{\frac{1}{2}}t'^{1/2}}
u(\xi,\tau ),\qquad \xi = s/t'^{1/2}.
\label{r_dyn}
\end{equation}
Namely, (\ref{Ricci}) admits a constant solution $u(\xi,\tau) = 1$,
and the linearisation around it gives the same linear operator
(\ref{ling}) as for mean curvature flow. Thus a pinching solution 
behaves as 
\begin{equation}
u(\xi,\tau )=1 + a(\tau) H_2(\xi/2) + o(\tau ^{-1}), 
\label{r_sol}
\end{equation}
where the equation for $a$ is $a_{\tau}=-8a^2$, with solution 
$a = 1/(8\tau)$. 

\subsubsection{Reaction-diffusion equations}
\label{sub:react}
The semilinear parabolic equation
\begin{equation}
u_{t}-\Delta u-\left\vert u\right\vert ^{p-1}u=0
\label{semilinear}
\end{equation}
is again closely related to the mean curvature flow problem (\ref{mc}).
Namely, disregarding the higher order term in $h_x$, (\ref{mc}) 
becomes 
\begin{equation}
h_t= h_{xx} - \frac{1}{h}.
\label{mc_simple}
\end{equation}
Putting $u = 1/h$ one finds 
\begin{equation}
u_t= u_{xx} + u^3 -2u_x^2/u,
\label{u_simple}
\end{equation}
which is (\ref{semilinear}) in one space dimension and $p=3$,
once more neglecting higher-order non-linearities. 
As before, (\ref{semilinear}) has the exact blow-up solution
\begin{equation}
u= (p-1)^{\frac{1}{1-p}}t'^{-\frac{1}{p-1}}. 
\label{semi_blow}
\end{equation}

If $1<p<p_{c}=\frac{d+2}{d-2}$, where $d$ is the space dimension, then there
are no other self-similar solutions to (\ref{semilinear}) \cite{GK85}, and
blow-up is of the form (\ref{semi_blow}) 
(see \cite{HV93}, \cite{MZ98} and \cite{GV02} for a
recent review). As in the case of mean curvature flow, corrections 
to (\ref{semi_blow}) are described by a slowly varying amplitude $a$:
\begin{equation}
u=t'^{1/(p-1)}(p-1)^{\frac{1}{1-p}}
\left[1 - a H_2(\xi/2) + O(1/\tau^2)\right], \quad \xi=x'/t'^{1/2},
\label{semi_corr}
\end{equation}
where $a$ obeys the equation 
\begin{equation}
a_{\tau} = -4p a^2.
\label{semi_slow}
\end{equation}
This result holds in 1 space dimension. In higher dimensions, one has 
to replace $x$ by the distance to the blow-up set.

This covers all range of exponents (larger than one, because otherwise there
is no blow-up) in dimensions $1$ and $2$. The situation if $p>p_{c}$ is not
so clear: if $p>1+\frac{2}{d}$ then there are solutions that blow-up and
"small" solutions that do not blow-up. Nevertheless, the construction of
solutions as perturbations of constant self-similar solutions holds for any $d$
and any $p>1$. A simple generalisation of (\ref{semilinear}) results from
considering a nonlinear diffusion operator,
\begin{equation}
u_t - \nabla\cdot(|u|^m\nabla u) = u^p
\label{semi_gen}
\end{equation}
and now the blow-up character depends on the two parameters m and p,
see \cite{Vazquez}.

\subsection{Cubic non-linearity: Cavity breakup and Chemotaxis}

More complex logarithmic corrections are possible if the linearisation around
the fixed point leads to a zero eigenvalue and cubic nonlinearities.

\subsubsection{Cavity break-up}
\label{cavity}

As shown in \cite{EFLS07}, the equation for a slender cavity or bubble is
\begin{equation}
\int_{-L}^L\frac{\ddot{a}(\xi,t)d\xi}{\sqrt{(x-\xi)^2+a(x,t)}} = 
\frac{\dot{a}^2}{2a} ,  \label{bernoulli}
\end{equation}
where $a(x,t)\equiv h^2(x,t)$ and $h(x,t)$ is the radius of the bubble.
Dots denote derivatives with respect to time $t$. 
The length $L$ measures the total size of the bubble. If
for the moment one disregards boundary conditions and looks for solutions to 
(\ref{bernoulli}) of cylindrical form, $a(x,t) = a_0(t)$, one can do the
integral to find
\begin{equation}
\ddot{a}_0\ln\left(\frac{4L^2}{a_0}\right) = \frac{\dot{a}_0^2}{2a_0}.
\label{cyl}
\end{equation}
It is easy to show that an an asymptotic solution of (\ref{cyl}) is given by
\begin{equation}
a_0 \propto \frac{t'}{\tau^{1/2}},  \label{cav0}
\end{equation}
corresponding to a power law with a small logarithmic correction. Indeed,
initial theories of bubble pinch-off \cite{LKL91,OP93} treated the case of
an approximately cylindrical cavity, which leads to the radial exponent $%
\alpha=1/2$, with logarithmic corrections.

\begin{figure}[tbp]
\centering
\includegraphics[width=8cm]{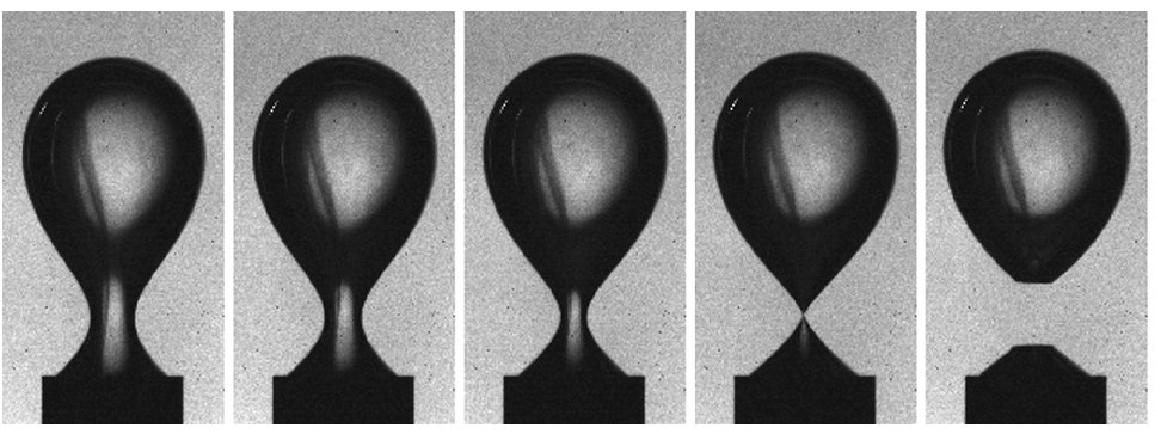}
\caption{ The pinch-off of an air bubble in water \protect\cite{TET07}. An
initially smooth shape develops a localised pinch-point. }
\label{shape}
\end{figure}

However both experiment \cite{TET07} and simulation \cite{EFLS07} show that
the cylindrical solution is unstable; rather, the pinch region is rather
localised, see Fig.~\ref{shape}. Therefore, it is not enough to treat the
width of the cavity as a constant $L$; the width $W$ is itself a
time-dependent quantity. In \cite{EFLS07} we show that to leading order the
time evolution of the integral equation (\ref{bernoulli}) can be reduced to
a set of ordinary differential equations for the minimum $a_0$ of $a(x,t)$,
as well as its curvature $a_0^{\prime \prime }$.

\begin{figure}[tbp]
\centering
\psfrag{t'}{\Large$t'$} \includegraphics[width=7cm]{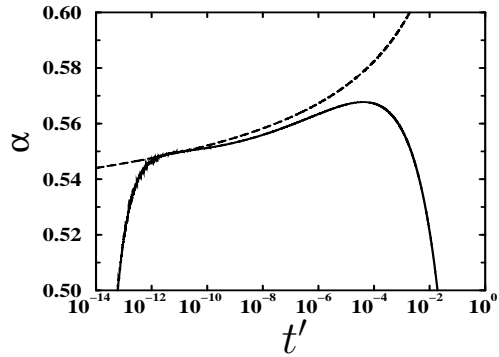}
\caption{A comparison of the exponent $\protect\alpha$ between full
numerical simulations of bubble pinch-off (solid line) and the leading order
asymptotic theory \eref{asymp-exp} (dashed line). }
\label{compare}
\end{figure}

Namely, the integral in (\ref{bernoulli}) is dominated by a local
contribution from the pinch region. To estimate this contribution, it is
sufficient to expand the profile around the minimum at $z=0$: $a(x,t) = a_0
+ (a^{\prime \prime }_0/2) z^2 + O(z^4)$. As in previous theories, the
integral depends logarithmically on $a$, but the axial length scale is
provided by the inverse curvature $W\equiv (2a_0/a^{\prime \prime
}_0)^{1/2}$. Thus evaluating (\ref{bernoulli}) at the minimum, one obtains
\cite{EFLS07} to leading order
\begin{equation}
\ddot{a}_0\ln(4W^2/a_0) = \dot{a}_0^2/(2a_0),  \label{a0}
\end{equation}
which is a coupled equation for $a_0$ and $W$. Thus, a second equation
is needed to close the system, which is obtained by evaluating the the
second derivative of (\ref{bernoulli}) at the pinch point:
\begin{equation}
\ddot{a}^{\prime \prime }_0\ln\left(\frac{8}{e^3a^{\prime \prime }_0}\right)
-2\frac{\ddot{a}_0a^{\prime \prime }_0}{a_0} = \frac{\dot{a}_0\dot{a}%
_0^{\prime \prime }}{a_0} - \frac{\dot{a}_0^2a_0^{\prime \prime }}{2a_0^2}.
\label{Deltaequ}
\end{equation}

The two coupled equations (\ref{a0}),(\ref{Deltaequ}) are most easily recast
in terms of the time-dependent exponents
\begin{equation}
2\alpha\equiv -\partial_{\tau}a_0/a_0, \quad 2\delta \equiv
-\partial_{\tau}a^{\prime \prime }_0/a^{\prime \prime }_0,  \label{exp}
\end{equation}
where $\beta = \alpha-\delta$, so $\alpha,\beta$ are generalisations of
the usual exponents in (\ref{ss}). The exponent 
$\delta$ characterises the time dependence of the aspect ratio $W$.
Returning to the collapse (\ref{cyl}) predicted for a constant solution, one
finds that $\alpha = 1/2$ and $\delta = 0$. In the spirit of the the
previous subsection, this is the fixed point corresponding to the
cylindrical solution. Now we expand the values of $\alpha$ and $\delta$
around their expected asymptotic values $1/2$ and $0$:
\begin{equation}
\alpha=1/2+u(\tau),\quad \delta =v(\tau).  \label{exp_ad}
\end{equation}
and put $w(\tau) = 1/\ln(a''_0)$. 

To leading order, the resulting equations are
\begin{equation}
u_{\tau} = u + w/4, \quad v_{\tau} = -v - w/4, 
\quad w_{\tau} = 2v w^2.
\label{l_sys}
\end{equation}
The linearisation around the fixed point thus has the eigenvalues
$0$ and $-1$, in addition to the eigenvalue $1$ coming from time 
translation. As before, the vanishing eigenvalue is the
origin of the slow approach to the fixed point observed for the 
present problem. The derivatives $u_{\tau}$ and $v_{\tau}$
are of lower order in the first two equations of (\ref{l_sys}), 
and thus to leading order $u=v$ and $v=-w/4$. Using this, the last equation
of (\ref{l_sys}) can be simplified to 
\begin{equation}
w_{\tau} = -w^3/2.
\label{deg_ad}
\end{equation}

Equation (\ref{deg_ad}) is analogous to \eref{a1},
but has a degeneracy of third order, rather than second order.
Equation \eref{deg_ad} yields, in an expansion for
small $\delta$ \cite{EFLS07},
\begin{equation}
\alpha=1/2+\frac{1}{4\sqrt{\tau}}+O(\tau), \quad \delta =\frac{1}{4\sqrt{\tau%
}} + O(\tau^{-3/2}).  \label{asymp-exp}
\end{equation}
Thus the exponents converge toward their asymptotic values $\alpha=\beta=1/2$
only very slowly, as illustrated in Fig.~\ref{compare}. This explains why
typical experimental values are found in the range $\alpha\approx 0.54-0.58$
\cite{TET07}, and why there is a weak dependence on initial conditions \cite%
{BMSSPL06}.

\subsubsection{Keller-Segel model for chemotaxis}
\label{sub:KS}
This model describes the aggregation of microorganisms driven by
chemotactic stimuli. The problem has biological meaning in 2 space
dimensions. If we describe the density of individuals by $u(x,t)$ and the
concentration of the chemotactic agent by $v(x,t)$, then the 
Keller-Segel system reads
\numparts
\begin{eqnarray}
u_{t} &=&\Delta u-\chi \nabla \cdot (u\nabla v), \label{KS1} \\
\Gamma v_{t} &=&\Delta v+(u-1), \label{KS2} 
\end{eqnarray}
\endnumparts
where $\Gamma $ and $\chi $ are positive constants. In \cite{HV96a,HV96b} 
it was shown that for radially symmetric solutions of (\ref{KS1}),(\ref{KS2})
singularities are such that to leading order $u$ blows up in the form of 
a delta function. The profile close to the singularity 
is self-similar and of the form
\begin{equation}
u(r,t)=\frac{1}{R^{2}(t)}U\left( \frac{r}{R(t)}\right),
\label{rad_ss}
\end{equation}
where 
\begin{equation}
R(t)=Ce^{-\frac{1}{2}\tau -\frac{\sqrt{2}}{2}\tau ^{\frac{1}{2}}-\frac{1}{4}%
\ln \tau +\frac{1}{4}\frac{\ln \tau }{\sqrt{\tau }}}(1+o(1))
\label{rad_R}
\end{equation}
and
\begin{equation}
U(\xi )=\frac{8}{\chi (1+\xi ^{2})}.
\label{rad_U}
\end{equation}

The result comes from a careful matched asymptotics analysis that, in our
notation, amounts to introducing the time-dependent exponent 
\begin{equation}
\gamma = -\partial _{\tau }R/R,
\label{flow_gamma}
\end{equation}
which has the fixed point $\gamma = 1/2$. Corrections are of the form
\begin{equation}
\gamma =\frac{1}{2}+\frac{\alpha }{2}\left( \alpha -\alpha ^{2}+1\right),
\label{gamma_eq}
\end{equation}
where $\alpha$ is controlled by a third-order non-linearity, 
as in the bubble problem:
\begin{equation}
\alpha_{\tau}=-\alpha ^{3}(1-\alpha+o(\alpha)).
\label{alpha_eq}
\end{equation}

\subsection{Beyond all orders: The nonlinear Schr\"{o}dinger equation}
\label{sub:NLSE}
The cubic nonlinear Schr\"{o}dinger equation
\begin{equation}
i\varphi _{t}+\Delta \varphi +\left\vert 
\varphi \right\vert ^{2}\varphi =0\,
\label{NLSE}
\end{equation}
appears in the description of beam focusing in a nonlinear optical medium,
for which the space dimension is $d=2$. Equation (\ref{NLSE}) belongs 
to the more general family of nonlinear Schr\"{o}dinger equations of the form
\begin{equation}
i\varphi _{t}+\Delta \varphi +\left\vert 
\varphi \right\vert ^{p}\varphi =0 , 
\label{NLSE_gen}
\end{equation}
and in any dimension $d$. Of particular interest, from the point of view 
of singularities, is the {\it critical case} $p=4/d$. In this case, 
singularities with slowly converging similarity exponents appear due 
to the presence of zero eigenvalues.
We will describe this situation below, based on the formal construction 
of Zakharov \cite{DNPZ}, later proved rigorously by Galina Perelman \cite{P01}.
At the moment, the explicit construction has only been given for $d=1$, 
that is, for the quintic Schr\"{o}dinger equation. The same blow-up
estimates have been shown to hold for any space dimension $d<6$ by 
Merle and Rapha\"el \cite{MR04}, \cite{MR06}, without making use of 
Zakharov's \cite{DNPZ} formal construction. Merle and Rapha\"el also show 
that the stable solutions to be described below are in fact global attractors. 

In the critical case (\ref{NLSE_gen}) becomes in d=1:
\begin{equation}
i\varphi _{t}+\varphi _{xx}+\left\vert \varphi \right\vert ^{4}\varphi =0.
\label{NLSE_quin}
\end{equation}
This equation has explicit self-similar solutions (in the sense that
rescaling $x\rightarrow \lambda x$, $t\rightarrow \lambda ^{2}t$, $\varphi
\rightarrow \lambda ^{\frac{1}{2}}\varphi $ leaves the solutions unchanged
except for the trivial phase factor $e^{-2i\mu _{0}\ln \lambda }$) of the
form
\begin{equation}
\varphi(x,t)=e^{i\mu_{0}\tau }e^{-\frac{\xi^{2}}{8}i}\frac{1}{t'^{
\frac{1}{4}}}\varphi_{0}(\xi ), \quad \xi=x'/t'^{1/2}.
\label{NLSE_ss}
\end{equation}
The function $\varphi _{0}(\xi )$ solves
\begin{equation}
-\varphi _{0,\xi\xi}+\varphi _{0}-\left\vert \varphi_{0}
\right\vert ^{4}\varphi _{0}=0,
\label{NLSE_phi0}
\end{equation}
and is given explicitly by
\begin{equation}
\varphi _{0}(\xi )=\frac{(3\mu _{0})^{\frac{1}{4}}}{\cosh ^{\frac{1}{2}}(2%
\sqrt{\mu _{0}}\xi )}.
\label{NLSE_osc}
\end{equation}

We seek solutions of (\ref{NLSE_quin}) using a generalisation of 
(\ref{NLSE_ss}), which allow for a variation of the phase factors,
and the amplitude to be different from a power law:
\begin{equation}
\varphi (x,t)=e^{i\mu (t)-i\beta (t)z^{2}/4}\lambda ^{\frac{1}{2}}(t)\varphi
_{a}(z),
\label{NLSE_ansatz}
\end{equation}
where $z=\lambda(t)x$ and $\varphi _{a}$ satisfies
\begin{equation}
-\varphi _{a,\xi\xi}+\varphi _{a}-\frac{1}{4}az^{2}\varphi
_{a}-\left\vert \varphi _{a}\right\vert ^{4}\varphi _{a}=0. 
\label{NLSE_a}
\end{equation}
When $h$ $(=\sqrt{a})$ is constant, (\ref{NLSE_ansatz}) is a solution 
of (\ref{NLSE_quin}) if $(\mu ,\lambda ,\beta )$ satisfy

\numparts
\begin{eqnarray}
\mu _{t} &=&\lambda ^{2} \label{NLSE1} \\
\lambda ^{-3}\lambda _{t} &=&\beta \label{NLSE2} \\
\beta _{t}+\lambda ^{2}\beta ^{2} &=&\lambda ^{2}h^{2}\label{NLSE3}.
\end{eqnarray}
\endnumparts

Notice that the equation for $\mu$ is uncoupled, so we only need to solve
the equations for $(\lambda,\beta)$ simultaneously and then integrate the
equation for $\mu$. It is interesting for the following that, in addition 
to the solutions for constant $a$, one can let $a$ vary slowly 
in time. The resulting system for $(\lambda,\beta,h)$ is
\numparts
\begin{eqnarray}
\lambda ^{-3}\lambda _{t} &=&\beta \label{slow1} \\
\beta _{t}+\lambda ^{2}\beta ^{2} &=&\lambda ^{2}h^{2} \label{slow2} \\
h_{t} &=&-c\lambda ^{2}e^{-S_{0}/h}/h \label{slow3}.
\end{eqnarray}
\endnumparts
Note the appearance of the factor $e^{-S_{0}/h}$ in the last equation,
which comes from a semiclassical limit of a linear Schr\"{o}dinger 
equation with appropriate potential (see \cite{P01}), and 
\begin{equation}
S_0= \int_0^2\sqrt{1-s^2/4}ds = \frac{\pi}{2}.
\label{circle}
\end{equation}
 $S_{0}$ is an 
It follows from the presence of this factor
that the non-linearity is beyond all orders, smaller than any given
power, in contrast to the examples given above. 

As in section \ref{cavity}, we rewrite the equations in terms of 
similarity exponents,
\begin{equation}
\alpha=-\frac{\lambda _{\tau }}{\lambda },\ 
\gamma = -\frac{\beta _{\tau }}{\beta }, \
\delta = -\frac{h_{\tau }}{h}
\label{NLSE_exp}
\end{equation}
to obtain the system:
\numparts
\begin{eqnarray} 
\alpha _{\tau } &=&-(1+2\alpha +\gamma )\alpha \label{NLSE_e1}\\
\gamma _{\tau } &=&(1+2\alpha +\gamma )\alpha -(\gamma +\alpha )(1+2\alpha
+2\delta -\gamma ) \label{NLSE_e2}\\
\delta _{\tau } &=&(-1-2\alpha +2\delta )\delta -\delta ^{2}\frac{S_{0}}{h}
\label{NLSE_e3} \\
h_{\tau } &=&-\delta h \label{NLSE_e4}. 
\end{eqnarray}
\endnumparts
The advantage of this formulation is that the exponents have fixed 
points. There are two families of equilibrium points for 
(\ref{NLSE_e1})-(\ref{NLSE_e4}):
\begin{enumerate}
\item[(1)] 
$\alpha =-\frac{1}{2},\ \gamma =0\ ,\delta =0,\ h$ 
arbitrary positive or zero.
\item[(2)] $\alpha =-1,\ \gamma =1\ ,\delta =0,\ h$ 
arbitrary positive or zero.
\end{enumerate}

We first investigate case (1) by writing
\begin{equation}
\alpha =-\frac{1}{2}+\alpha _{1},\ \gamma =\gamma _{1},\
\delta =\delta _{1},\ h =h_{1}. 
\label{NLSE_lin}
\end{equation}
The final fixed point corresponding to the singularity is 
going to be $\alpha_1=\gamma_1=\delta_1=h_1=0$. However, there
are also equilibrium points for {\it any} $h>0$, in which case 
the linearisation reads:
\numparts
\begin{eqnarray}
\alpha _{1,\tau } &=&\alpha _{1}+\frac{1}{2}\gamma _{1} \label{NLSE_lin1}\\
\gamma _{1,\tau } &=&-\gamma _{1}+\delta _{1} \label{NLSE_lin2}\\
\delta _{1,\tau } &=&2\delta _{1}^{2}-2\alpha _{1}\delta _{1}-\delta _{1}^{2}
\frac{S_{0}}{h}\label{NLSE_lin3}. 
\end{eqnarray}
\endnumparts
This system has the matrix
\[
A=\left(
\begin{array}{ccc}
1 & \frac{1}{2} & 0 \\
0 & -1 & 1 \\
0 & 0 & 0
\end{array}
\right),
\]
whose eigenvalues are: $1,0$, and $-1$. The vanishing eigenvalue 
corresponds to the line of equilibrium points for $h>0$, 
the positive eigenvalue to the direction of instability generated by
a change in blow-up time. The eigenvector corresponding to the 
negative eigenvalue gives the direction of the stable manifold.

At the point $h=0$, there is an additional vanishing eigenvalue,
and the equations become:
\numparts
\begin{eqnarray}
\alpha _{1,\tau ^{\prime }} &=&(\alpha _{1}+\frac{1}{2}\gamma _{1})h_{1}
\label{h01} \\
\gamma _{1,\tau ^{\prime }} &=&(-\gamma _{1}+\delta _{1})h_{1} \label{h02}\\
\delta _{1,\tau ^{\prime }} &=&(2\delta _{1}^{2}-2\alpha _{1}\delta
_{1})h_{1}-\delta _{1}^{2}S_{0} \label{h03} \\
h_{1,\tau ^{\prime }} &=&-\delta _{1}h_{1}^{2} \label{h04}, 
\end{eqnarray}
\endnumparts
where $d\tau ^{\prime }=d\tau /h_{1}$. The first two equations reduce
to leading order to $\gamma_1 = \delta_1 h_1$ and 
$\alpha_1 = -\delta_1 h_1^2/2$, while the last two equations reduce to
the nonlinear system:
\begin{equation}
\delta _{1,\tau ^{\prime }} = -\delta _{1}^{2}S_{0}, \quad
h_{1,\tau ^{\prime }} = -\delta _{1}h_{1}^{2}, \quad
\tau_{\tau'}=h_1.
\label{NLSE_los}
\end{equation}
In the original $\tau$-variable, the dynamical system is 
\begin{equation}
\delta _{1,\tau} = -\delta _{1}^{2}S_0/h_1 \quad
h_{1,\tau ^{\prime }} = -\delta _{1}h_{1}, 
\label{NLSE_orig}
\end{equation}
which controls the approach to the fixed point. The system (\ref{NLSE_orig})
is two-dimensional, corresponding to the two vanishing eigenvalues. 

Integrating the first equation of (\ref{NLSE_los}) one gets
$\delta_1\sim 1/(S_0\tau')$, and thus using the second equation 
$h_1 \sim S_0/\ln \tau'$. From the last equation one obtains 
to leading order $\tau' \sim \tau\ln\tau/S_0$, so that
\begin{equation}
h_{1}\sim \frac{S_{0}}{\ln\tau}\ ,\ \delta _{1}\sim \frac{1}{\tau \ln\tau}.
\label{h1d1}
\end{equation}
Thus we can conclude that
\begin{equation}
\alpha (\tau ) \simeq \frac{1}{2}-\frac{1}{2\tau \ln\tau},\quad
\gamma (\tau ) \simeq \frac{1}{\tau \ln\tau},\quad
\delta (\tau ) \simeq \frac{1}{\tau \ln\tau}.
\label{NLSE_concl}
\end{equation}
In this fashion, one can construct a singular solution such that
\begin{eqnarray}
\varphi (x,t)=e^{-i\tau\ln \tau-i\frac{1}{t'}%
x^{2}/4}\frac{(\ln \tau)^{\frac{1}{4}}}{t'^{\frac{1}{4}}%
}\varphi _{h^{2}\tau}\left( \frac{(\ln \tau)^{\frac{1}{2}}}{%
t'^{\frac{1}{2}}}x\right) \nonumber \\
\sim e^{-i\tau\ln \tau}\frac{(\ln \tau
)^{\frac{1}{4}}}{t'^{\frac{1}{4}}}\varphi _{0}\left( \frac{%
(\ln \tau)^{\frac{1}{2}}}{t'^{\frac{1}{2}}}x\right)
\label{fashion}
\end{eqnarray}
Note the remarkable smallness of this correction to the ``natural'' 
scaling exponent of $t'^{1/4}$, which enters only 
as the logarithm of logarithmic time $\tau$. 

The fixed points (2) can be analysed in a similar fashion. 
The linearisation leads to
\numparts
\begin{eqnarray}
\alpha _{1,\tau } &=&2\alpha _{1}+\gamma _{1} \label{unst1} \\
\gamma _{1,\tau } &=&\gamma _{1} \label{unst2} \\
\delta _{1,\tau } &=&\delta _{1} \label{unst3}.
\end{eqnarray}
\endnumparts
All eigenvalues are positive, so one cannot expect
these equilibrium points to be stable.

One may also consider the blow-up of vortex solutions to both
critical and supercritical solutions to nonlinear Schrödinger 
equation in 2D. These are a subset of the general solutions to 
NLSE that present a phase singularity at a given point. The singularities
appear in the form of collapse of rings at that point. Both the
existence of such solutions and their stability have been considered 
recently in \cite{FGW07,FG08}.

\subsubsection{Other nonlinear dispersive equations}
\label{other_disp}

The nonlinear Schr\"{o}dinger equation belongs to the broader 
class of nonlinear dispersive equations, for which many questions 
concerning existence and
qualitative properties of singular solutions are still open.
Nevertheless, there have been recent developments that we describe
next.

The Korteweg-de Vries (KdV) equation
\begin{equation}
u_{t}+(u_{xx}+u^{2})_{x}=0
\label{kdv}
\end{equation}
describes the propagation of waves with large wave-length in a dispersive
medium. For example, this is the case of water waves in the shallow 
water approximation, where $u$ represents the height of the wave. In
the case of an arbitrary exponent of the nonlinearity, (\ref{kdv})
becomes the generalised Korteweg de Vries equation:
\begin{equation}
u_{t}+(u_{xx}+u^{p})_{x}=0\ ,\ p>1 .
\label{kdv_gen}
\end{equation}

Based on numerical simulations, \cite{BDKM95} conjectured the existence of
singular solutions of (\ref{kdv_gen}) with type-I self-similarity 
if $p\geq5 $. 
In \cite{MM02a}, \cite{MM02b} it was shown that in the 
{\it critical case} $p=5$ solutions may blow-up both in finite and in
infinite time.
Lower bounds on the blow-up rate were obtained, but they exclude blow-up in
the self-similar manner proposed by \cite{BDKM95}.

The Camassa-Holm equation
\begin{equation}
u_{t}-u_{xxt}+3u_{x}u=2u_{x}u_{xx}+u_{xxx}u
\label{CH}
\end{equation}
also represents unidirectional propagation of surface waves on a shallow
layer of water. It's main advantage with respect to KdV is the existence of
singularities representing breaking waves \cite{CE98}. The structure of
these singularities in terms of similarity variables has not been addressed
to our knowledge.

\section{Travelling wave}
\label{travel}
The pinching of a liquid thread in the presence of an 
external fluid is described by the Stokes equation \cite{LS98}. For 
simplicity, we consider the case that the viscosity $\eta$ of the 
fluid in the drop and that of the external fluid are the same. 
An experimental photograph of this situation is shown in 
Fig.~\ref{lava}. 
To further simplify the problem, we make the assumption 
(the full problem is completely analogous) that the fluid
thread is slender. Then the equations given in \cite{CBEN99} simplify to
\begin{equation}
\label{interface}
h_t = -v_x h/2 - v h_x,
\end{equation}
where 
\begin{equation}
\label{velocity}
v = \frac{1}{4} \int_{x_-}^{x_+} 
\left(\frac{h^2(y)}{\sqrt{h^2(y)+(x-y)^2}} \right)_y \kappa \;dy,
\end{equation}
and the mean curvature is given by (\ref{mean}). 
Here we have written the velocity in units of the capillary
speed $v_{\eta} = \gamma/\eta$. The limits of integration 
$x_-$ and $x_+$ are for example the positions of the plates
which hold a liquid bridge \cite{P43}.

Dimensionally, one would once more expect a local solution of the
form
\begin{equation}\label{56}
    h(x,t)= t'H\left(\frac{x'}{t'}\right),
\nonumber
\end{equation}
and $H(\xi)$ has to be a linear function at infinity to match
to a time-independent  outer solution. In similarity variables, 
\eref{velocity} has the form
\begin{equation}\label{58}
    V(\xi)=\frac
    14\int^{x_{b/t'}}_
      {-x_b/t'}\left(\frac{H^2(\eta)}
    {\sqrt{H^2(\eta)+(\xi-\eta)^2}})\right)_{\eta}\kappa \;d\eta .
\end{equation}
We have chosen $x_b$ as a real-space variable close to the pinch-point,
such that the similarity description is valid in $[-x_b,x_b]$. 
But if $H$ is linear, the integral in (\ref{58}) diverges
like $b\ln t'$, where 
\begin{equation}\label{63}
  b=-\frac14\left[\frac{H_+}{1+H_+^2}+
\frac{H_-}{1+H_-^2}\right]. 
\end{equation}
Here $H_+$ and $H_-$ are the slopes of the similarity profile
at $\pm\infty$. But this means that a simple ``fixed point'' 
solution (\ref{56}) is impossible. 

However by subtracting the singularity as $t'\rightarrow 0$, 
one can define a self-similar velocity profile according to 
\begin{equation}\label{64}
   V^{\rm{(fin)}}(\xi)=\lim_{\Lambda\to \infty}\frac 14
   \int^\Lambda_{-\Lambda}\left(\frac{H^2(\eta)}
{\sqrt{H^2(\eta)+(\xi-\eta)^2}}\right)_{\eta} \kappa \;d\eta
 + b\ln \Lambda,
\end{equation}
where now 
\begin{equation}
\label{v_rel}
V(\xi) = V^{\rm{(fin)}}(\xi) - b\tau,
\end{equation}
and an arbitrary constant has been absorbed into $V^{\rm{(fin)}}$.
In terms of $V^{\rm{(fin)}}$, and putting 
\begin{equation}\label{self_time}
    h(x,t)= t'H \left(\xi,\tau\right),
\nonumber
\end{equation}
the dynamical system for $H$ becomes 
\begin{equation}\label{travel_equ}
 H_{\tau} = H - \left(\xi  + V^{\rm (fin)}\right)H_{\xi} 
 - HV^{\rm (fin)}_{\xi}/2 + b\tau H_{\xi}. \nonumber 
\end{equation}
This equation has a solution in the form of a travelling wave:
\begin{equation}\label{travel_ansatz}
    H(\xi,\tau) = \overline{H}(\zeta), \quad 
    V^{\rm (fin)}(\xi,\tau) = \overline{V}(\zeta), \quad \mbox{where}
\quad \zeta = \xi - b\tau. 
\end{equation}
The profiles $\overline{H},\overline{V}$ of the travelling wave 
obey the equation 
\begin{equation}\label{61}
  \overline{H} - (\zeta+\overline{V})\overline{H}_{\zeta}
 = \overline{H}\;\overline{V}_{\zeta}/2. \nonumber 
\end{equation}

The numerical solution of the integro-differential equation 
(\ref{61}) gives
\begin{equation}\label{66}
    h_{\min}=a_{\rm{out}}v_\eta t', \quad \mbox{where} \quad
    a_{\rm{out}}=0.033. 
\end{equation}
The slope of the solution away from the pinch-point are
given by 
\begin{equation}\label{slope}
 H_+=6.6 \quad \mbox{and} \quad  H_-=-0.074,
\end{equation}
which means the solution is very asymmetric, as confirmed
directly from Fig.~\ref{lava}. These results are reasonably
close to the exact result, based on a full solution of the 
Stokes equation \cite{CBEN99}; in particular, the normalised 
minimum radius is $a_{\rm{out}}=0.0335$ for the full problem. 

\section{Limit cycles}
\label{limit} 
An example for this kind of blow-up was introduced into the
literature in \cite{C93} in the context of cosmology. There is considerable
numerical evidence \cite{Gund03} that discrete self-similarity occurs at the
mass threshold for the formation of a black hole. The same type of
self-similarity has also been proposed for singularities of the Euler
equation \cite{PSS92,PS05}, the porous medium equation driven by 
buoyancy \cite{PSS92}, and for a variety of other 
phenomena \cite{Sor98}. A reformulation of the original cosmological 
problem leads to the following system:
\numparts
\begin{eqnarray}  
&& f_x=\frac{(a^2-1)f}{x},  \label{cosm:a} \\
&& (a^{-2})_x=\frac{1-(1+U^2+V^2)/a^2}{x},  \label{cosm:b} \\
&& (a^{-2})_t=\left[\frac{(f+x)U^2-(f-x)V^2}{x}+1\right]/a^2-1,
\label{cosm:c} \\
&& U_x=\frac{f[(1-a^2)U+V]-xU_t}{x(f+x)},  \label{cosm:d} \\
&& V_x=\frac{f[(1-a^2)U+V]+xV_t}{x(f-x)}.  \label{cosm:e}
\end{eqnarray}
\endnumparts
In \cite{MG03}, the self-similar description corresponding to the system 
(\ref{cosm:a})-(\ref{cosm:e}) 
was solved using formal asymptotics and numerical shooting
procedures. This leads to the solutions observed in \cite{C93}. We now
propose another system, which shares some of the structure of 
(\ref{cosm:a})-(\ref{cosm:e}), but which we are able to solve analytically:
\numparts
\begin{eqnarray} 
&& u_t(x,t) = 2f(x,t)v(x,t),  \label{f:a} \\
&& v_t(x,t) = -2f(x,t)u(x,t),  \label{f:b} \\
&& f_t(x,t) = f^2(x,t).  \label{f:c}
\end{eqnarray}
\endnumparts
The system (\ref{f:a})-(\ref{f:c}) is driven by the simplest type of 
blow-up equation (\ref{f:c}), and can be solved using characteristics.
However, in the spirit of this review, we transform to similarity 
variables according to:
\numparts
\begin{eqnarray}
&& u = U(\xi,\tau) \label{circ:a} \\
&& v = V(\xi,\tau) \label{circ:b} \\
&& f = t'^{-1}F(\xi,\tau) \label{circ:c} 
\end{eqnarray}
\endnumparts
It is seen directly from (\ref{f:c}) that $f$ first blows up at a local
maximum $f_{max}>0$. Near a maximum, the horizontal scale is the 
square root of the vertical scale $t'$, and thus we must
have $\xi = x'/t'^{1/2}$. With that, the similarity equations
become
\numparts
\begin{eqnarray}  \label{simc:all}
&& U_{\tau} = -\xi U_{\xi}/2 + FV \label{simc:a} \\
&& V_{\tau} = -\xi V_{\xi}/2 - FU \label{simc:b} \\
&& F_{\tau} = -F - \xi F_{\xi}/2 + F^2. \label{simc:c} 
\end{eqnarray}
\endnumparts

The fixed point solution of the last equation is 
\begin{equation}
F = \frac{1}{1 + c\xi^2},  \label{1st_circ}
\end{equation}
where $c>0$ is a constant. The equations for $U,V$ are solved by the
ansatz
\begin{equation}
U = U_0\sin\left(C(\xi) + \tau \right),\quad  
V = U_0\cos\left(C(\xi) + \tau \right),\quad  
\label{ansatz_circ}
\end{equation}
and for the function $C(\xi)$ one finds 
\begin{equation}
\xi C'(\xi)/2 = F-1, 
\label{c_circ}
\end{equation}
with solution $C(\xi) = -\ln(1+c\xi^2)$. 
Thus (a single component of) the singular solution is indeed of 
the general form 
\begin{equation}
U = \psi(\phi (\xi )+\tau ),  \label{dss_structure}
\end{equation}
where $\psi $ is {\it periodic} in $\tau $. This is a particularly 
simple version of discretely self-similar behaviour, i.e. when $T$ is 
the period of $\psi$, the same self-similar picture is obtained for
$\tau=\tau_0 + nT$. 

\section{Strange attractors and exotic behaviour}
\label{sec:strange}

In connection to limit cycles and in the context of singularities in
relativity, a few interesting situations have been found numerically quite
recently. One of them is the existence of Hopf bifurcations where a
self-similar solution (a stable fixed point) is transformed into a discrete
self-similar solution (limit cycle) as a certain parameter varies 
(see \cite{HE97}). Other kinds of bifurcations, for example of the 
Shilnikov type, are found as well \cite{ABT06}. Before coming to simple 
explicit examples, we mention that possible complex dynamics in $\tau$ 
has long been suggested for simplified versions of the inviscid Euler 
equations \cite{PS92,PS92b,PSS92}. For a critical discussion of this
work, see \cite{ES94,MB02}.

The problems considered in these papers were the 2D axisymmetric Euler 
equations with swirl, which produces a centripetal force. In the limit
that the rotation is confined to a small annulus, the direction of
acceleration is locally uniform, and the equation reduces to that of
2D Boussinesq convection, where the centripetal force is replaced by 
a ``gravity'' force. Another related model is 2D porous medium convection, for
which the equation reads 
\begin{equation}
\frac{\partial T}{\partial t} + \left(T{\bf e}_y - \nabla \phi\right)\cdot
\nabla T = 0,
\label{porous}
\end{equation}
where ${\bf v} = T{\bf e}_y - \nabla \phi$ plays the role of the velocity
field and $T$ is the temperature. The potential $\phi$ follows from
the constraint of incompressibility, which gives $\triangle\phi = T_y$. 
Simulations provide evidence of a self-similar dynamics of the form 
\cite{PSS92} 
\begin{equation}
T = t'^{\eta} M({\bf x}'/t'^{1+\eta},\tau),
\label{porous_dyn}
\end{equation}
where $\eta$ is approximately 0.1 and $M$ is a function that 
is slowly varying with $\tau$. 

Depending on the model, both periodic behaviour as well as more 
complicated, chaotic motion has been observed in numerical 
simulations. Oscillations of
temperature in $\tau$ are motivated by the observation that a 
sharp, curved interface (i.e. the transition region between a 
rising ``bubble'' of hot fluid and its surroundings) 
becomes unstable and rolls up. However, owing to 
incompressibility, the sheet is also stretched, which stabilises 
the interface, leading to an eventual decrease in gradients. Locality
suggests that this process could repeat itself periodically on smaller 
and smaller scales \cite{PSS92}. However, simulations of the Euler
equation have also shown examples of a more complicated dependence 
on $\tau$, which might be chaotic behaviour \cite{PS92}. We also
mention that corresponding chaotic behaviour has been proposed for 
the description of spin glasses in the theory of critical phenomena
\cite{MBK82}. We now give some explicit examples of chaos in the 
description of a singularity.

In section \ref{quad} we treated a system of an infinite number of ordinary
differential equations for the coefficients of the expansion of an arbitrary
perturbation to an explicit solution. Such high-dimensional systems in
principle allow for a rich variety of dynamical behaviours, including those
found in classical finite dimensional dynamical systems, such as chaos.
Consider for instance an equation for the perturbation $g$ (the analogue of (%
\ref{gequ})) of the form
\begin{equation}
g_{\tau }=\mathit{L}g+F(g,g),  \label{ggen}
\end{equation}
where $\mathit{L}g$ is a linear operator. Assuming an appropriate non-linear
structure for the function $F$, an arbitrary nonlinear (chaotic) dynamics
can be added.

To give an explicit example of a system of PDE's exhibiting chaotic dynamics,
consider the structure of the example given in section \ref{limit}. It can be
generalised to produce \textit{any} low-dimensional dynamics near the
singularity, as follows by considering the system (\ref{f:a})-(\ref{f:c}) 
\numparts
\begin{eqnarray}  
&& u^{(i)}_t(x,t) = 2fF_i(\{u^{(i)}\}), \quad i=1,\dots,n,
\label{ch:a} \\
&& f_t(x,t) = f^2(x,t).  \label{ch:d}
\end{eqnarray}
\endnumparts
Using the ansatz analogous to (\ref{ansatz_circ}):
\begin{equation}
u^{(i)} = U^{(i)}\left(C(\xi) + \tau,\xi\right),
\label{ansatz_lor}
\end{equation}
and choosing $C(\xi) = -\ln(1+c\xi^2)$, one obtains the system
\begin{equation}
U^{(i)}_{\tau} = F_i\left\{U^{(i)}\right\}.
\label{res_lor}
\end{equation}

To be specific, we consider $n=3$ and
\begin{equation}
F_1 = \sigma(u^{(2)}-u^{(1)}), \quad F_2 = \rho
u^{(1)}-u^{(2)}-u^{(1)}u^{(3)}, \quad F_3 = u^{(1)}u^{(2)}-\beta u^{(3)},
\label{sys_L}
\end{equation}
so that (\ref{res_lor}) becomes the Lorenz system \cite{Strogatz94}.
As before, for $t^{\prime }\rightarrow 0$, the variable $\tau$ goes to
infinity, and near the singularity one is exploring the long-time behaviour
of the dynamical system (\ref{ansatz_lor}). 
In the case of (\ref{sys_L}), and for
sufficiently large $\rho$, the resulting dynamics will be chaotic.
Specifically, taking $\sigma=10$, $\rho=28$, and $\beta=8/3$, as done by
Lorenz \cite{L63}, the maximal Lyapunov exponent is $0.906$. 
The initial conditions with which (\ref{ansatz_lor}) is to be solved
depend on $\xi$. Thus the chaotic dynamics will follow a completely 
different trajectory for each space point. As a result, it will be 
very difficult to detect self-similar behaviour of this type as such,
even if data arbitrarily close to the singularity time is taken. 
If for example a rescaled spatial picture is observed at constant
intervals of logarithmic time $\tau$, the spatial structure of the singularity
will appear to be very different. However, as pointed out in \cite{PS92},
chaotic motion is characterised by unstable periodic orbits, for 
which one could search numerically. 

\section{Multiple singularities}
\label{multiple}
The singularities described so far occur at a single point $x_0$ at a given 
time $t_{0}$. This need
not be the case, but blow-up may instead occur on sets of varying complexity,
including sets of finite measure. We begin with a case where singularity
formation involves two different points in space. 

\subsection{Hele-Shaw equation}
\label{HS_sub}
A particularly rich singularity structure is found for a
special case of (\ref{thin_film}) in one space dimension
with $n=1$. Dropping the second term on the right, which
will typically be small, one arrives at 
\begin{equation}
h_{t} + (hh_{xxx})_x = 0.
\label{HS}
\end{equation}
This is a simplified model for a neck of liquid of width $h$ confined 
between two parallel plates, a so-called Hele-Shaw cell. 
which is a simplified model for the free surface in a so-called
Hele-Shaw cell \cite{A96}. Breakup of a fluid neck inside the 
cell corresponds to $h$ going to zero in finite time. 

Singular solutions displaying type-I self-similarity 
would be of the form
\begin{equation}
h(x,t) = t'^{\alpha}H(x'/t'^{(\alpha+1)/4}),
\label{HS_ss}
\end{equation}
but are never observed. Instead, several types of pinch solutions different 
from (\ref{HS_ss}) have been found for (\ref{HS}) using 
a combination of numerics and asymptotic arguments 
\cite{CDGKSZ93,DGKZ93,ABB96}. On one hand, singularities 
exhibit type-II self-similarity. On the other hand, the simple 
structure (\ref{HS_ss}) is broken by the fact that the location of 
the pinch point is {\it moving} in space. The root for this 
behaviour lies in the fact that two singularities are {\it interacting}
over a distance much larger than their own spatial extend. 
Below we report on three different kinds of singularities whose
existence has been confirmed by numerical simulation of (\ref{HS}). 

The first kind of singularity was called the {\it imploding 
singularity} in \cite{ABB96}, since it consists of two self-similar
solutions which form mirror images, and which collide at the singular
time. Locally, the solution can be written
\begin{equation}
h(x,t) = t'^6H((x'+at')/t'^3),
\label{HS_imp}
\end{equation}
where $-a$ is the constant speed of the singular point. Note that 
the scaling exponents do not agree with (\ref{HS_ss}). The reason is that 
the singularity is moving, so $h$ is the solution of 
\begin{equation}
hh_{xxx} = J(t')\equiv t'^3,
\label{HS_move}
\end{equation}
where $J$ is determined by matching to an outer region. 
The similarity profile $H$ is a solution of the equation 
$HH''' = 1$, with boundary conditions 
\begin{equation}
H(\eta)\propto \eta^2/2,\; \eta \rightarrow -\infty; \quad
H(\eta)\propto \sqrt{8/3}(A-\eta)^{3/2},\; \eta \rightarrow \infty.
\label{HS_asymp}
\end{equation}

\begin{figure}[t]
\centering
\includegraphics[width=0.4\hsize]{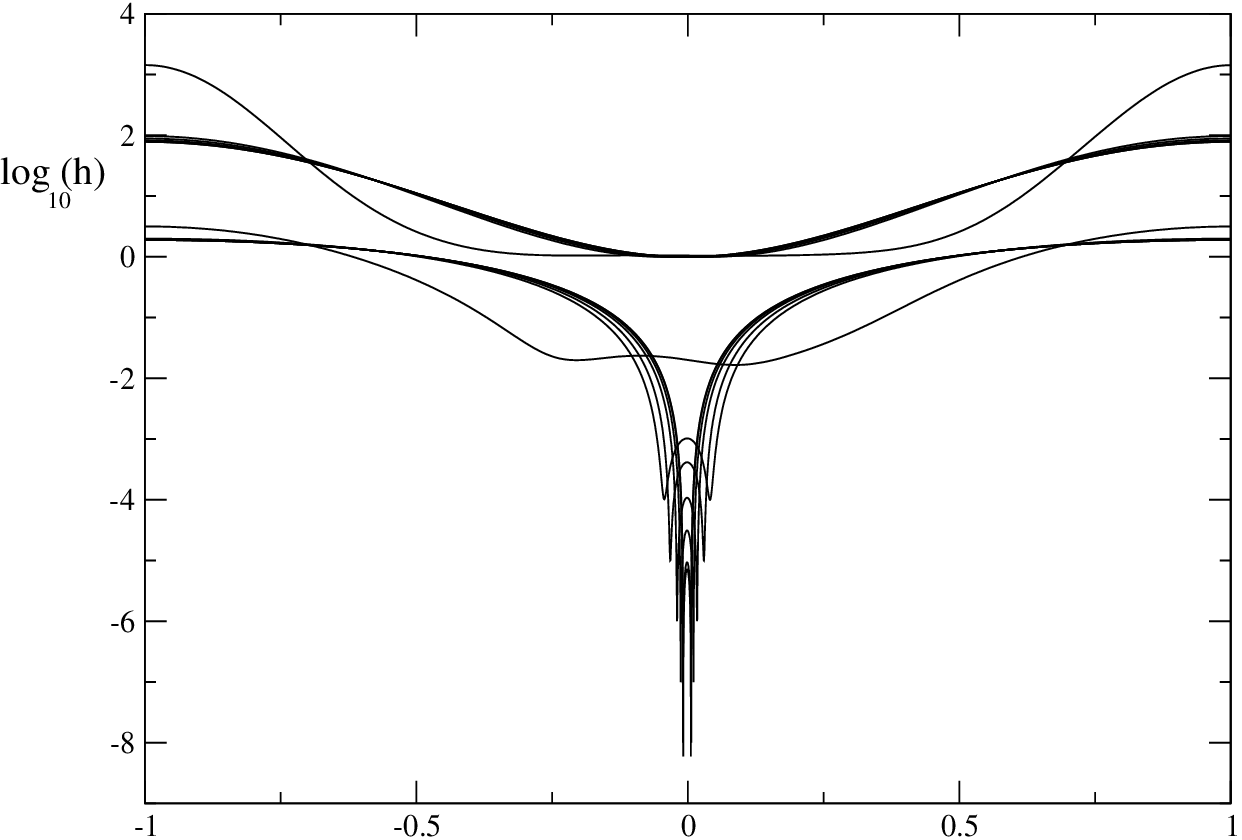}
\caption{A simulation of (\ref{HS}) with spatially periodic boundary 
conditions and initial condition (\ref{HS_init}), with $w=0.02$ and
$\delta=0.1$. \label{HS_simulation}}
\end{figure}

One might wonder whether this behaviour is generic, in the sense that 
it might depend on the initial conditions being exactly symmetric 
around the eventual point of blow up. The simulation of (\ref{HS}) 
shown in Fig.~\ref{HS_simulation} shows that this is {\it not} the case. 
The initial condition is 
\begin{equation}
h(x,0) = 1 - (1-w)\left[\frac{3}{2}\cos\pi x - \frac{6}{10}\cos2\pi x 
+ \frac{1}{10}\cos3\pi x (1+\delta\sin2\pi x) \right],
\label{HS_init}
\end{equation}
which for $\delta = 0$ reduces to the symmetric initial condition 
considered by \cite{ABB96}. The type of singularity that is observed
(or no singularity at all) depends on the parameter $w$. The simulation
shown in Fig.~\ref{HS_simulation} shows that even at finite $\delta$
(non-symmetric initial conditions) the final collapse is described 
by a symmetric solution. 

\begin{figure}[t]
\centering
\includegraphics[width=0.4\hsize]{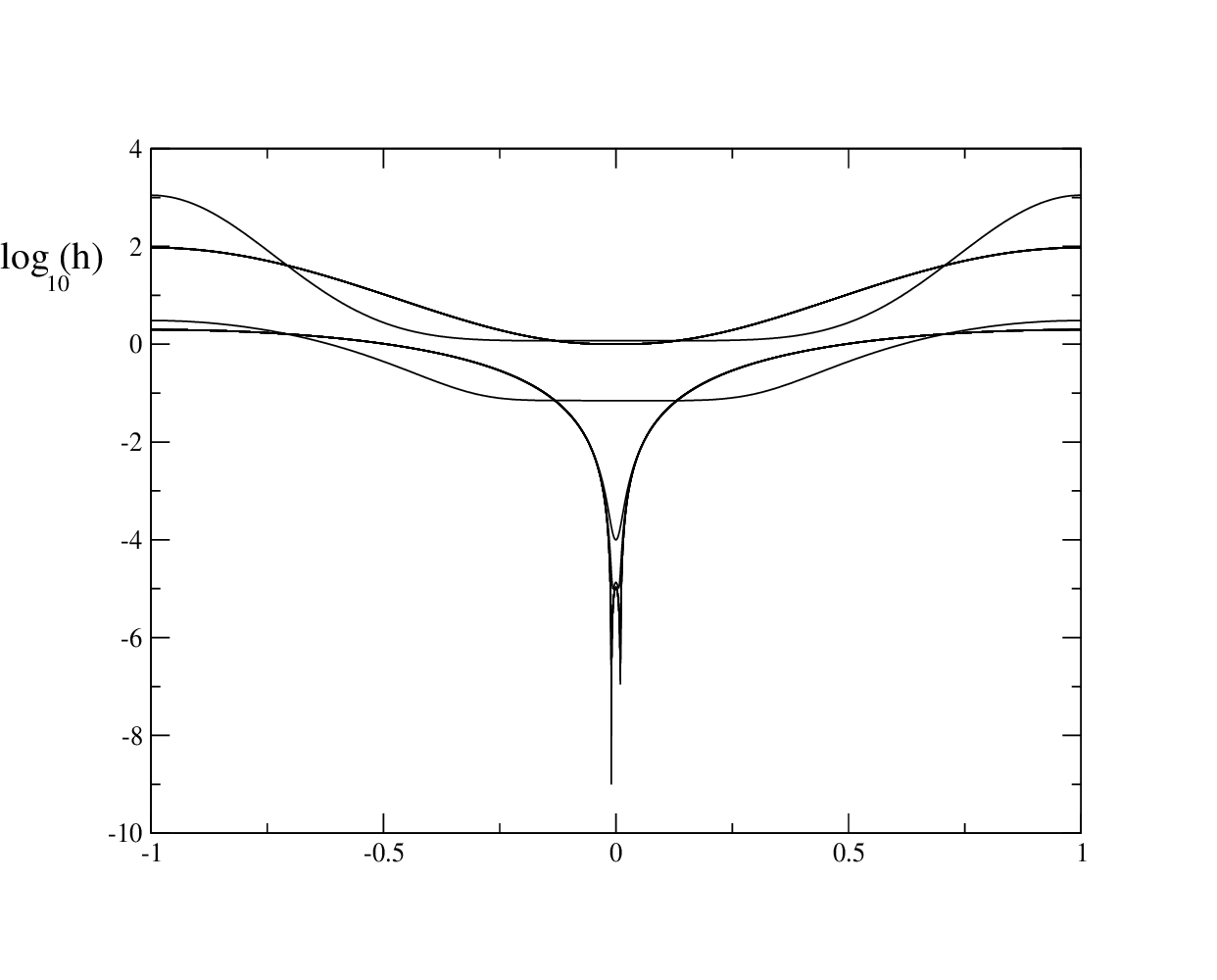}
\caption{Same as Fig.\ref{HS_simulation}, but both parameters $w=0.07$ and
$\delta=0.01$. \label{HS_simulation2}}
\end{figure}

The second kind is the {\it exploding singularity} \cite{ABB96}, 
since now the two self-similar solutions are moving apart, 
cf. Fig.\ref{HS_simulation2}. This time even a very small asymmetry
($\delta=1/100$) makes one pinching event ``win'' over the other. 
However, this does not affect the asymptotics described briefly below. 
Locally, the solution can be written
\begin{equation}
h(x,t) = \delta^2(t')H((x'-at')/\delta(t')),
\label{HS_exp}
\end{equation}
with $\delta = t'/ln(t')$, which is similar to examples considered 
in section \ref{sec:centre}. However, an additional 
complication consists in the fact that the singularity is moving,
so there is a coupling to the parabolic region between 
the two pinch-points. This matching is unaffected by the fact 
that in the simulation shown in Fig.~\ref{HS_simulation2} 
one side of the solution touches down first. 
In \cite{ABB96}, a possible generalisation is also conjectured,
which has the form 
\begin{equation}
h(x,t) = \delta^2(t')H((x'-at'^{\frac{r-1}{2}})/\delta(t')),
\label{HS_gen}
\end{equation}
and $\delta = t'^{\frac{r-1}{2}}/\ln t'$. In principle, any
value of $r$ is possible, but numerical evidence has been found
for $r\approx 3$ (above) and $r\approx 5/2$ only. 

Finally, a third type is the {\it symmetric singularity} 
of \cite{ABB96}, which does not move. In that case, the structure 
of the solution is  
\begin{equation}
h(x,t) = h_0(t')H((x'/\delta(t')),
\label{HS_symm}
\end{equation}
with $h_0 = \delta^2 P(\ln\delta)$, where $P$ is a polynomial.
The time dependence of $\delta$ is not reported. Evidently, many
aspects of the exploding and of the symmetric singularity remain 
to be confirmed and/or to be worked out in more detail. 

The most intriguing feature of the Hele-Shaw equation (\ref{HS}) is 
that several types of {\it stable} singularities have been observed 
for the same equation. Within a one-parameter family of smooth
initial conditions, all three types of singularities can be realized
as $h\rightarrow 0$. Each type is observed over an interval of the 
parameter $w$. Near the boundary of the intervals, a very interesting
crossover phenomenon occurs: the solution is seen to follow one
type of singularity at first (the exploding singularity, say), and
then crosses over to a solution of another singularity (the
imploding singularity). The dynamics of each singularity can be
followed numerically over many decades in $t'$. By tuning $w$,
the crossover can be made to occur at arbitrarily small values of $h$. 

The switch in behaviour is driven by the slow dynamics of scaling
regions exterior to (\ref{HS_imp}) or (\ref{HS_exp}). It is a
signature of the very long-ranged interactions (both in real space
as well as in scale), that exist in (\ref{HS}). Thus an outside
development can trigger a change of behaviour that is taking place
on the local scale of the singularity. To mention another example,
applying different boundary conditions for the pressure at the outside
of the cell can change the singular behaviour completely \cite{Bertozzi}. 
This makes the crossover behaviour of (\ref{HS}) very different from 
that observed for drop pinch-off (cf. (\ref{hequ}),(\ref{vequ})),
which is driven by a change in the dominant balance between different
terms in (\ref{vequ}). 

\subsection{Semilinear wave equation}
\label{sub:semi}
It appears that the Hele-Shaw equation is not an isolated example, 
but rather is representative of a more general phenomenon. Namely,
another example of a potentially complex singularity structure is 
the semilinear wave equation
\begin{equation}
u_{tt}-\Delta u=|u|^{p-1} u ,\ p>1.
\label{semi-wave}
\end{equation}
It has trivial singular solutions of the form 
\begin{equation}
u(x,t)=b_{0}(T-t)^{-\frac{2}{p-1}}, 
\label{semi_sol}
\end{equation}
with $b_{0}=\left[ \frac{2(p+1)}{(p-1)^{2}}\right] ^{\frac{1}{p-1}}$.
Nevertheless, the existence of different self-similar solutions is known in a
few particular cases, like the case $p\geq 7$, where $p$ is an odd integer 
(see \cite{BMW07}) or in space dimension $d=1$ (see \cite{MZ07}). 

The character of the blow-up is controlled by the blow-up curve $T(x)$,
which is the locus where the equation first blows up at a given point
in space. It has been shown for $d=1$ \cite{MZ08} that there exists a
set of {\it characteristic} points, where the blow-up curve locally 
coincides with the characteristics of (\ref{semi-wave}). The set of 
non-characteristic points $I_0$ is open, and $T$ is $C^1$ on $I_0$. 
Recently, it has been shown \cite{MZ08_prep} that the blow-up at 
characteristic points is of type II. Even more intriguingly,
it appears \cite{MZ08_prep} that the structure of blow-up
at these points is such that the singularity results from the 
{\it collision} of two peaks at the blow-up point, very similar 
to the observation shown in Fig.~\ref{HS_simulation}. 

\subsection{More complicated sets}
In the Hele-Shaw equation of the previous subsection, different parts 
of the solution, characterised by different scaling laws, interacted with
each other. In the generic case, however, finally blow-up only
occurred at a single point in space. An example where singularities 
may even occur on sets of finite measure is given by reaction-diffusion
equations of the family
\begin{equation}
u_{t}-\Delta u=u^{p}-b\left\vert \nabla u\right\vert^{q}\quad \mbox{for}
\quad x \in\Omega . \label{react-diff}
\end{equation}
where $\Omega$ is any bounded, open set in dimension $d$.
Depending on the values of $p>1$ and $q>1$ singularities of (\ref{react-diff})
may be regional ($u$ blows up in subsets of $\Omega $ of
finite measure), or even global (the solution blows-up in the whole domain);
see for instance \cite{Souplet} and references therein.

Singularities may even happen in sets of fractional Hausdorff dimension,
i.e., fractals. This is the case of the inviscid one-dimensional system for
jet breakup (cf. \cite{FV00}) and might be case of the Navier-Stokes system in
three dimensions, where the dimension of the singular set at the time of
first blow-up is at most $1$ (cf. \cite{CKN82}). This connects to the
second issue we did not address here. It is the nature of the singular sets
both in space and time, i.e. including possible continuation of solutions
after the singularity. In some instances, existence of global in time
(for all $0\leq t<\infty $) solutions to nonlinear problems can be
established in a \textit{weak sense}. For example, this has been 
achieved for systems like the Navier Stokes equations \cite{CF89}, 
reaction-diffusion equations \cite{Smoller}, and hyperbolic systems of 
conservation laws \cite{Dafermos}. Weak solutions allow for singularities
to develop both in space and time. In the case of the three-dimensional 
Navier-Stokes system, the impossibility of singularities "moving" in time,
that is of curves $\mathbf{x}=\mathbf{\varphi }(t)$ within the singular set
is well-known \cite{CKN82}. Hence, provided certain kinds of singularities do
not persist in time, the question is how to continue the solutions after a
singularity has developed.

\ack
A first version of this paper was an outgrowth of discussions between 
the authors and R. Deegan, preparing a workshop on singularities at the 
Isaac Newton Institute, Cambridge. The present version was written during
the programme: ``Singularities in mechanics: formation, propagation and 
microscopic description'', organised with C. Josserand and L. Saint-Raymond,
which took place between January and April 
2008 at the Institut Henri Poincar\'e in Paris. We are grateful to all
participants for their input, in particular C. Bardos, M. Brenner, 
M. Escobedo, F. Merle, H. K. Moffatt, Y. Pomeau, A. Pumir, J. Rauch,
S. Rica, L. Vega, T. Witten, and S. Wu. We also thank J. M. Martin-Garcia and 
J. J. L. Velazquez for fruitful discussions and for providing us with 
valuable references. 

\section*{References}
\providecommand{\newblock}{}


\end{document}